\documentclass[onecolumn]{rsauthor}

\usepackage{graphicx}
\usepackage{amsmath}
\usepackage{amssymb}
\usepackage{amsfonts}
\usepackage{amsthm}
\usepackage{verbatim}

\usepackage{xspace}
\newcommand{\ket}[1]{\ensuremath{|#1\rangle}\xspace}
\newcommand{\bra}[1]{\ensuremath{\langle #1|}\xspace}
\newcommand{\psh}[2]{\ensuremath{\langle #1|#2\rangle}\xspace}
\allowdisplaybreaks[4]

\jname{Philosophical Transactions of the Royal Society A}
\begin{document}

\title{Calculation of multifractal dimensions in spin chains}
\author{Y. Y. Atas and E. Bogomolny}
\address{Univ. Paris-Sud, CNRS, LPTMS, UMR8626,  91405 Orsay, France}


\abstract{It was demonstrated in Ref.~\cite{ab}, [Phys. Rev. E \textbf{86}, 021104,  (2012)],  that the ground-state wave functions for a large variety of one-dimensional spin-$\frac{1}{2}$ models are multifractals in the natural spin-$z$ basis.  We present here the  details of analytical derivations and numerical confirmations of these results.}

\maketitle

\section{Introduction}

Many-body quantum systems is an everlasting subject of theoretical physics. Today increase of computer power makes it possible to treat problems with a few tenths of particles  numerically which opens new possibilities of their investigation. 

Here we consider one of the oldest many-body interacting models, namely one-dimensional spin chains.  The archetypical example  is the 
XYZ Heisenberg model \cite{heisenberg}  for $N$ spins-$\tfrac{1}{2}$  in external fields  
\begin{equation}
\mathcal{H}=-\sum_{n=1}^N\Big[\frac{1+\gamma}{2}\sigma_n^{x}\sigma_{n+1}^x  +\frac{1-\gamma}{2} \sigma_n^{y}\sigma_{n+1}^y 
+\frac{\Delta}{2}  \sigma_n^{z}\sigma_{n+1}^z +\lambda \sigma_n^z+\alpha \sigma_n^{x}\Big]
\label{general_H}
\end{equation} 
and its various specifications for different values of parameters. $\sigma_n^{x,y,z}$  are the usual Pauli matrices at site $n$.

Many different methods were developed to find the spectra of such Hamiltonians \cite{bethe}-\cite{mattis}. The calculation of wave functions is more difficult even for integrable models. In the natural basis of $z$-component of each spin, $| \vec{\sigma} \rangle =| \sigma_{1}\cdots \sigma_{N} \rangle$ with $\sigma_j=\pm 1$ any wave function of spin-$\tfrac{1}{2}$ can be represented as a formal sum
\begin{equation}
\mathbf{\Psi}=\sum_{\{\vec{\sigma}\}} \Psi_{\vec{\sigma}} |\vec{\sigma}\rangle
\label{psi_expansion}
\end{equation}
where the summation is taken over all $M=2^N$ configurations with $N$ being the total number of spins. 

In such basis, Hamiltonians like \eqref{general_H} are represented by $M\times M$ matrix.  In general, the coefficients $\Psi_{\vec{\sigma}}$ are obtained by matrix diagonalization. As the size of the Hilbert space grows exponentially with the number of spins, exact diagonalization quickly becomes intractable and an iterative method of diagonalization is eventually necessary. Even in integrable cases  wave functions of spin chains still require the knowledge of exponentially large number of coefficients which looks quite erratic (see figures below) and their structure is not well understood. 

In Ref.~\cite{ab} it was shown that the ground state (GS) wave functions for spin chains are multifractals in the spin-$z$ basis.  The multifractality as usual (see e.g. \cite{mandelbrot,kadanoff}) means that moments of wave functions scale non-trivially with the Hilbert space dimension $M$
\begin{equation}
P_q\equiv \sum_{\{\vec{\sigma}\}}|\Psi_{\vec{\sigma}} |^{2q}\underset{M\to\infty}{\sim}M^{-\tau(q)}
\label{tau_q}
\end{equation}
and $\tau(q)=D_q(q-1)$ where $D_q$ are called fractal dimensions.

More precisely,  let $S_R(q,M)$ be the R\'enyi entropy \cite{renyi} for an eigenfunction \eqref{psi_expansion} of a $M\times M$ matrix 
\begin{equation}
S_R(q,M)=-\frac{1}{q-1}\ln \Big (\sum_{\{\vec{\sigma}\}}|\Psi_{\vec{\sigma}} |^{2q}\Big )
\label{renyi_entropy}
\end{equation}  
with   normalized coefficients  $\Psi_{\vec{\sigma}}$, $\sum_{\{\vec{\sigma}\}}|\Psi_{\vec{\sigma}} |^{2}=1$. 

Fractal dimensions, $D_q$, are determined  from the behaviour of the R\'enyi entropy \eqref{renyi_entropy} in the limit $M\to\infty$ \cite{mirlin}
\begin{equation}
D_q=\lim_{M\to \infty} \frac{S_R(q,M)}{\ln M}\ .
\label{fractal}
\end{equation}
If this limit is zero,  $D_q=0$,  only a small number of coefficients $\Psi_{\vec{\sigma}}$ gives large contribution and one refers to such functions as localized functions. In the opposite situation, when almost all coefficients  $\Psi_{\vec{\sigma}}$ are of the same order, 
$|\Psi_{\vec{\sigma}}|^2\approx 1/M$, it is plain that $D_q=1$. Such functions are called delocalized ones. (For $d$-dimensional systems fully delocalized functions have $D_q=d$.) In general case,  $D_q$ has a non-linear dependence of $q$ and corresponding functions are labelled as multifractals.  

Multifractality is a general notion introduced  to characterize quantitatively strong and irregular fluctuations of  various quantities  \cite{mandelbrot}-\cite{stanley}. It appears in very different physical contexts from turbulence \cite{Meneveau} to human heartbeat dynamics \cite{Ivanov}.  Wave function multifractality had attracted a wide attention when it was recognized that it appears at special points in some condensed matter problems (e.g. at the point of metal-insulator transition in the $3$-dimensional Anderson model \cite{mit}  and in $2$-dimensional quantum Hall effect) (cf.  \cite{mirlin}). More simple critical random matrix models whose eigenfunctions have multifractal properties consist on matrices whose off-diagonal elements decrease as the first power of the distance from the diagonal \cite{levitov,seligman}: 
$M_{i j}\sim |i-j|^{-1}$ when $|i-j|\gg 1$. 

In all such critical models, multifractality is a non-trivial consequence of the concurrence between delocalization due to spreading and the localization due to randomness. In spin models \eqref{general_H} there is no random parameters and the fact that the GS wave functions are multifractal may seem strange. Nevertheless, by combining numerical and analytical calculations it  has been  proved  in \cite{ab}.  

The notion of multifractality, similar to the localization, depends on the basis. A function may be localized in one basis and delocalized in another. Spin chains are defined in spin basis (cf. \eqref{general_H}) and we investigate multifractal properties only in this basis though for certain problems the use of another basis, e.g. the  fermion one \cite{lieb}, may be useful.      

The purpose of this paper is twofold. First, we present certain details of analytical  calculations in \cite{ab}. Second, by the mere definition \eqref{fractal} the fractal dimensions are defined by the limiting procedure. As there is practically no cases where $D_q$ are known analytically, their careful determination is sensitive to the method of interpolation used to calculate the limit $N\to\infty$ from the data with relatively small $N$. This question is important in the applications but rarely discuss in the literature. We collect data obtained by the direct diagonalization of Hamiltonian matrices for models with up to 13 spins and the ones from iterative Lanczos algorithm \cite{Lanczos} up to 19 spins and compare different methods of interpolation.

The plan of this paper is the following. We start in Section~\ref{sec_binomial} with informal introduction to the multifractal formalism based on an example of the binomial measure. Though it is the simplest mathematical example of multifractality, it appears that there exits particular cases of spin chains \eqref{general_H} where the GS has  exactly the same structure as the binomial measure. In Section~\ref{numerics} two standard numerical methods of finding GS functions, the direct diagonalization and the Lanczos method,  are  briefly discussed. Section~\ref{QIM} is devoted to the investigation of quantum Ising model in a transverse field  which is one of the most studied spin chain model. By combining analytical and numerical calculations, we prove  that its ground state wave functions is multifractal. A generalization of the Ising model, namely, the  XY model is discussed in Section~\ref{XY}. Similar to the Ising model, this model is also integrable which permits to find analytically certain fractal dimensions. Special attention is given to the so-called factorising field were the XY model has exact and  simple factorising GS wave function. It is in this case that GS wave function can be described by the binomial cascade, thus proving rigorously the  multifractality of  this wave function.  In Section~\ref{XXZ} properties of GS wave functions for the XXZ and XYZ models are briefly discussed.  Section~\ref{conclusion} concludes the paper. For clarity we choose the parameters such that all terms in the Hamiltonian are non-positive. Due to the Perron-Frobenius theorem it implies that expansion coefficients of the  ground state wave function can be chosen non-negative. For the QIM and the XY models we impose that their ground states are  ferromagnetic but for the XXZ and XYZ models we choose them anti-ferromagnetic.

\section{Binomial measure}\label{sec_binomial}

To get an insight to the multifractality we consider first the simplest example of the 
multifractal measures built by iterating a  procedure called a (binary) multiplicative cascade \cite{mandelbrot}.

\begin{figure}[!t]
\begin{center}
\includegraphics[ width=.7\linewidth,clip ]{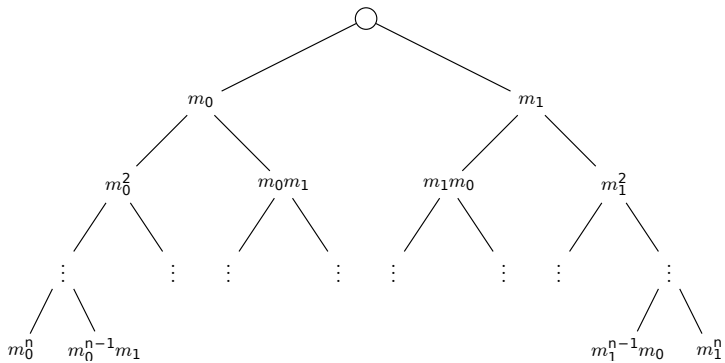}
\end{center}
\caption{Tree representation of the binomial cascade. The sum of the elements over a line is equal to one.}
\label{fig_tree}
\end{figure}

In the first step of the cascade, the unit interval is divided in two equal subinterval and one associates a mass $m_{0}$ (a measure) to the left subinterval $[0,1/2]$ and a mass $m_{1}$ to the right subinterval $[1/2,1]$. The two positive numbers $m_{0,1}$ are such that $m_{0}+m_{1}=1$ (convenient parametrization is $m_{0}=\cos^2 \theta$, $m_{1}=\sin^2 \theta$).  At stage 2 of the iteration we apply the same procedure to the two previous subinterval and the measures associated with four subinterval are: 
\begin{eqnarray}
\mu_{2}\left([0,1/4]\right)&=&m_{0}m_{0}, \qquad \mu_{2}\left([1/4,1/2]\right)=m_{0}m_{1},\nonumber\\ 
\mu_{2}\left([1/2,3/4]\right)&=&m_{1}m_{0}, \qquad \mu_{2}\left([3/4,1]\right)=m_{1}m_{1}.
\end{eqnarray}
This process can be visualized graphically in a binary tree (see Fig.~\ref{fig_tree}).

After $N$ iterations, one gets $M=2^N$ subintervals of the form 
\begin{equation}
\left[\frac{k}{2^N},\frac{k+1}{2^N}\right],\qquad k=0,1,\cdots,2^{N}-1.
\label{intervals}
\end{equation}
Let us associate a variable $\sigma_j=-1$ for the branch $j$ on the binary tree when it turns to the left and   $\sigma_j=1$ if it turns to the right.  Now  each interval  \eqref{intervals} corresponds uniquely to $|\vec{\sigma}\rangle = | \sigma_{1}\cdots \sigma_{N} \rangle$  with $\sigma_j=\pm 1$ and the  measure of  such interval is 
\begin{equation}
\mu_{N}\left(\left[\frac{k}{2^N},\frac{k+1}{2^N}\right]\right)\equiv p_{\vec{\sigma}} =m_{0}^{\varphi(\,\vec{\sigma}\, )}m_{1}^{N-\varphi(\, \vec{\sigma}\, )}, 
\label{measure_binomial}
\end{equation}
where $\varphi(\,\vec{\sigma}\,)$ is the number of left moves i.e. the number of $-1$ in the vector  $\vec{\sigma}$.

\begin{figure}[!ht,clip]
\begin{minipage}{.3\linewidth}
\begin{center}
\includegraphics[angle=-90, width=.99\linewidth,clip ]{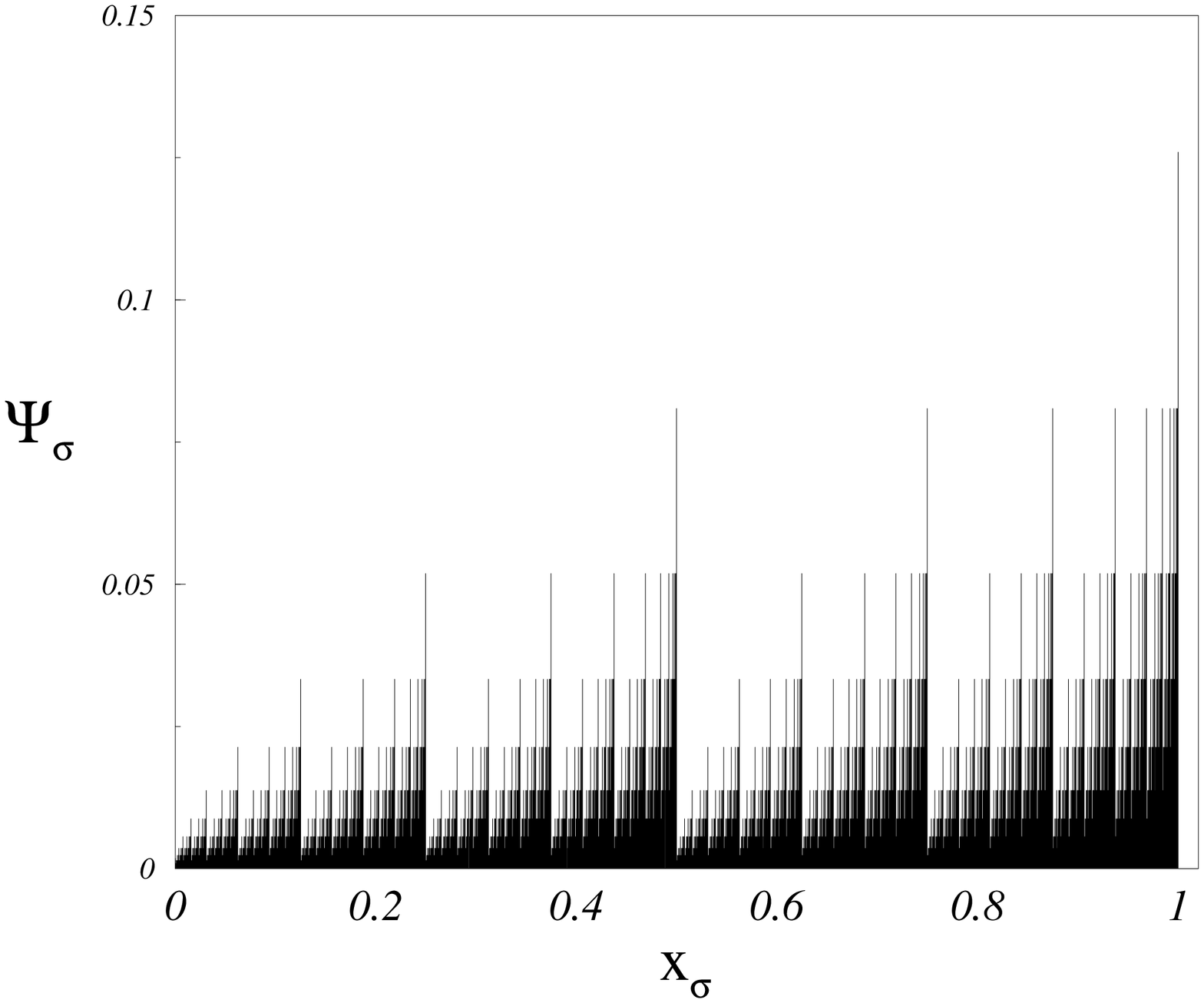}\\
 (a)
\end{center}
\end{minipage}
\begin{minipage}{.3\linewidth}
\begin{center}
\includegraphics[angle=-90, width=.99\linewidth,clip ]{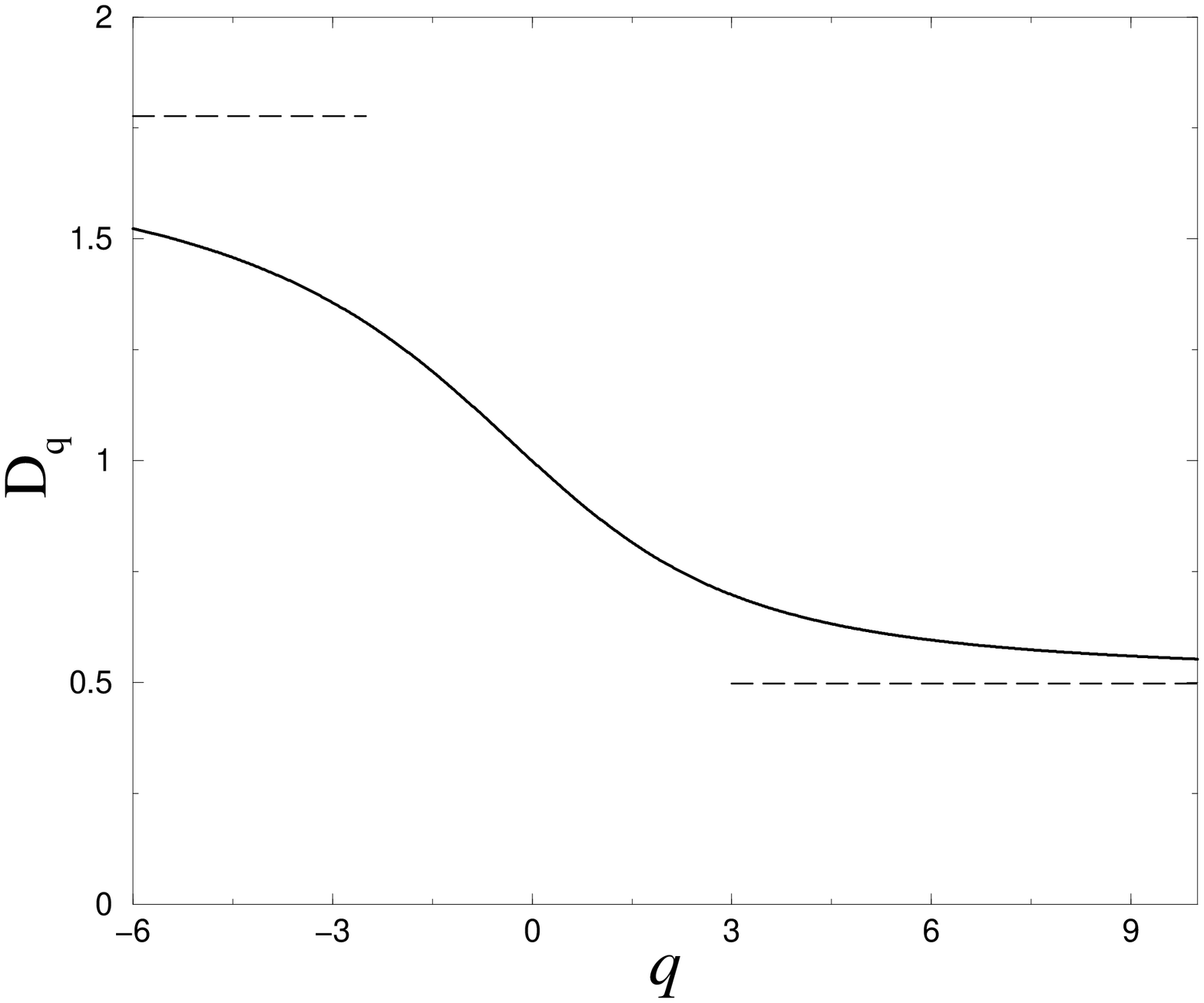}\\ 
 (b)
\end{center} 
\end{minipage}
\begin{minipage}{.3\linewidth}
\begin{center}
\includegraphics[angle=-90, width=.99\linewidth,clip ]{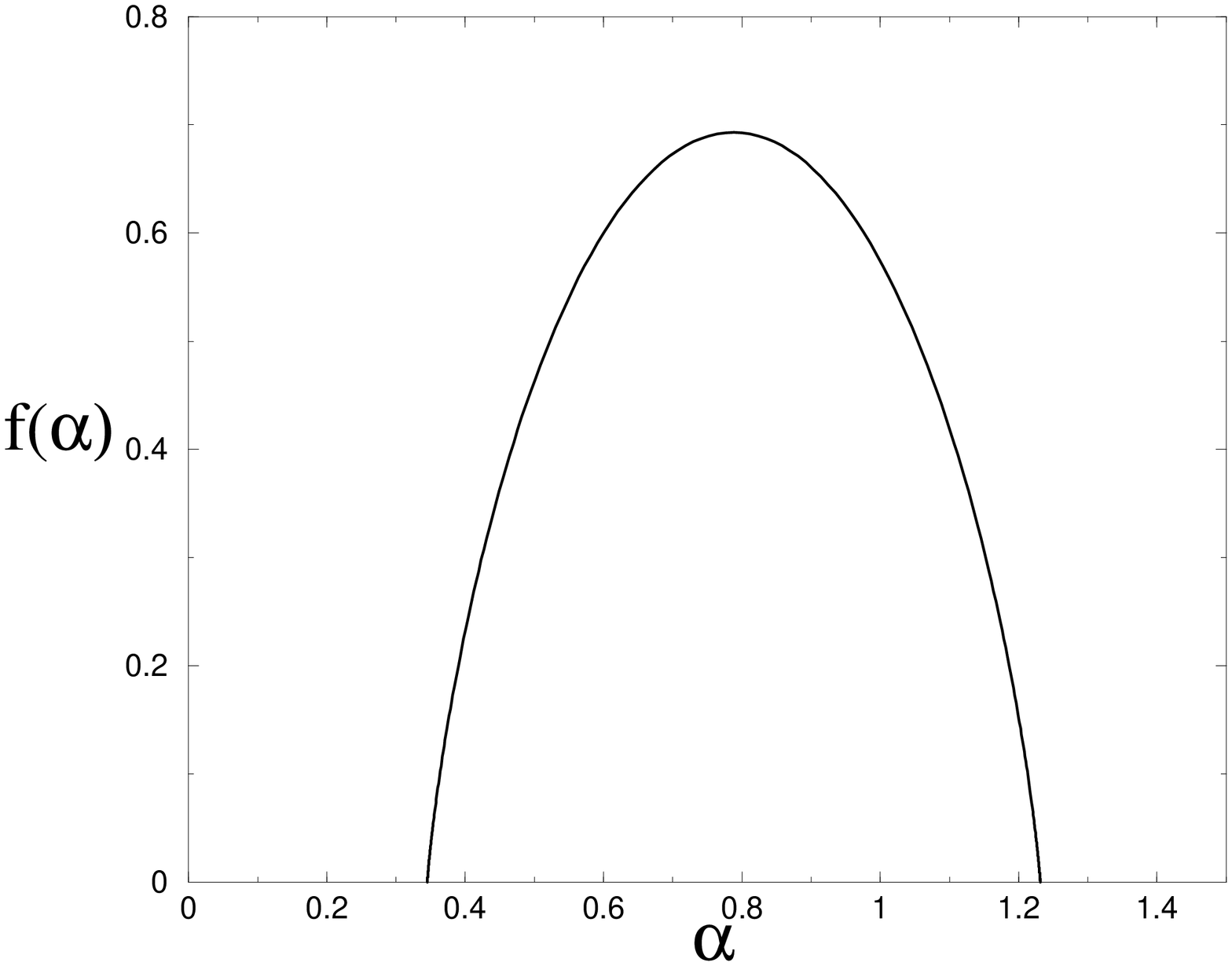} \\
(c)
\end{center}
\end{minipage}
\caption{(a) Binomial measure for $N=12$ steps of the binomial cascade for $m_{0}=\cos^2 1 \approx 0.292$ and  $m_{1}=\sin^2 1\approx 0.708$.  (b) Fractal dimensions for the same measure.  Dashed line indicate the limiting values $D_{\pm \infty}$ \eqref{limiting_values}. 
(c) Singularity spectrum for the  the same measure. }
\label{fig_binomial}
\end{figure}

Iteration of this procedure generate an infinite sequence of measure which converge to the binomial measure:
\begin{equation}
\mu =\lim_{N \to \infty } \mu_{N}.
\end{equation}
Fig.~\ref{fig_binomial} a) illustrates the function $\Psi_{\vec{\sigma}}\equiv \sqrt{p_{\vec{\sigma}}}$ obtained after $12$ iterations for $m_{0}=\cos^2 1$. The abscissa $x_{\vec{\sigma}}$  (with $0\leq x_{\vec{\sigma}}\leq 1$) is related to a binary code of binomial cascade as follows
\begin{equation}
 x_{\vec{\sigma}}=(2^{N}-1)^{-1}\sum_{n=1}^N 2^{n-2}(1+\sigma_n).
 \label{binary_code}
\end{equation}

Since the mass is conserved during the cascade, i.e. the sum of  elements  along a horizontal  line of the tree equals $1$   (due to the binomial theorem, hence the name), the measure  (\ref{measure_binomial}) is a probability measure.
\subsection{Fractal dimensions}

Let $S_R(q,M)$ be the R\'enyi entropy \eqref{renyi_entropy} for a discrete normalized probability distribution $\lbrace p_{\vec{\sigma}} \rbrace$   after $N$ steps ($M=2^N$)
\begin{equation}
S_R(q,M)=-\frac{1}{q-1}\ln \Big (\sum_{\{\vec{\sigma}\}} p_{\vec{\sigma}}^{q}\Big ).
\end{equation} 
For the binomial measure one gets
\begin{equation}
p_{\vec{\sigma}}\equiv |\Psi_{\vec{\sigma}}|^2=m_0^nm_1^{N-n},
\label{proba_binomial}
\end{equation}
where $n$ is the number of left moves along the binary tree (or number of $-1$ in sequence $\vec{\sigma}$). Therefore
\begin{equation}
\sum_{\{\vec{\sigma}\}}p_{\vec{\sigma}}^{q}=\sum_{j=0}^{N} C_N^j m_0^{q j }m_1^{q (N-j)}=(m_0^q+m_1^q)^N\, .
\end{equation}
Here $C_N^j $ are the usual binomial coefficients.

Fractal dimensions, $D_q$, are defined  from the limiting behaviour of the R\'enyi entropy  in the limit $M\to\infty$ \eqref{fractal}. 
For the binomial measure 
\begin{equation}
D_{q}=\frac{\ln (m_{0}^q+m_{1}^q)}{(1-q)\ln 2}. 
\label{binomial_dimension}
\end{equation}
The form of this function is shown in the Fig. \ref{fig_binomial} b). In particular, for large $q$ fractal dimensions tend to the limits
\begin{equation}
D_{\infty}=-\frac{\ln (\mathrm{max}(m_{0},m_{1}))}{\ln 2}, \qquad D_{-\infty}=-\frac{\ln (\mathrm{min}(m_{0},m_{1}))}{\ln 2}\ .
\label{limiting_values}
\end{equation}
As we shall see below, the curves of $D_{q}$ as a function of $q$ for all spin chain models considered in the paper have the same characteristic shape.
\subsection{Singularity spectrum} 

The example of the binomial measure is also convenient for informal discussion of other quantities of interest in multifractal formalism \cite{mandelbrot,kadanoff}, \cite{mirlin}. The most important of them is the singularity spectrum, commonly denoted by $f(\alpha)$.  To clarify physical interpretation of formulae below, it is  useful (but, of course,  not necessary) to define fictitious "energy levels", $E_{\vec{\sigma}}$   as follows
\begin{equation}
|\Psi_{\vec{\sigma}} |^2= \mathrm{e}^{-E_{\vec{\sigma}}},\qquad E_{\vec{\sigma}}=-\ln |\Psi_{\vec{\sigma}} |^2\ .
\label{thermodynamics}
\end{equation}      
Then expression \eqref{tau_q} takes the form
\begin{equation}
Z_q\equiv \sum_{\{\vec{\sigma}\}}|\Psi_{\vec{\sigma}} |^{2q}=\sum_{\{\vec{\sigma}\}}\mathrm{e}^{-q E_{\vec{\sigma}}}
\end{equation}
This sum can physically be interpreted as the canonical partition function for a system with total energies $E_{\vec{\sigma}} $ and the temperature $1/q$. As it is usual in thermodynamics one may introduce the density of energy levels as 
\begin{equation}
\rho(E)=\mathrm{e}^{S(E)}
\end{equation}
where $S(E)$ is the entropy. Then the previous sum reduces to
\begin{equation}
Z_q=\int \mathrm{e}^{S(E)-qE}\mathrm{d}E\ .
\label{Z_q}
\end{equation}
By analogy with thermodynamics it is natural to assume that the entropy $S(E)$ as well as the total energy $E$ are extensive  functions of number of particles $N$. It leads to the following expression for the entropy  
\begin{equation}
S(E)=Nf(E/N)
\label{extensive}
\end{equation}
with a certain function $f(\alpha)$.  (In application to fractal dimensions one has to insist that $N=\ln M$ where $M$ is the dimension of the Hilbert space. It introduces a constant factor between $N$ and the number of spins and will result in redefinition of $f(\alpha)$. We shall implicitly take into account this fact.)

Using \eqref{extensive}, the integral in \eqref{Z_q} at large $N$ can be calculated by the saddle point method,  and it is plain that 
\begin{equation}
Z_q \sim \mathrm{e}^{-N \tau(q)},\qquad \tau(q)=q\alpha -f(\alpha), \qquad \alpha= \tau^{\prime}(q)\ . 
\end{equation}
For the binomial cascade $\tau(q)$ is known (cf. \eqref{binomial_dimension}), $\tau(q)=-\ln (m_{0}^q+m_{1}^q)/\ln 2$  and $f(\alpha)$ can implicitly be calculated by the Legendre transform
\begin{equation}
f(\alpha)=q\alpha-\tau(q), \qquad \alpha= \tau^{\prime}(q)\ .
\end{equation}
In Fig.~\ref{fig_binomial} c) the singularity spectrum $f(\alpha)$  for the binomial cascade with $m_0=\cos^2 (1)\approx  0.292$ is presented. For other models $f(\alpha)$ has a similar form \cite{mirlin}. 

In the above thermodynamic language $\tau(q)$ and $f(\alpha)$ play the role, respectively, of the free energy and the entropy par particle. Of course, the multifractal formalism itself \cite{mandelbrot,kadanoff} does not require references to thermodynamics. But we think that the latter, being well known, sheds a particular light to the notion of multifractality and  clarifies the use and the meaning of different quantities.

In principle,  $D_q$ and $f(\alpha)$ contain the same information but in this paper we shall focus on the calculation of  fractal dimensions only.


\section{Numerical approaches}\label{numerics}

For completeness we briefly discuss main numerical methods of calculation of the ground state wave function for spin models like in  \eqref{general_H}.
A basis state of the chain will be noted $\ket{\vec{\sigma}}$ with 
\begin{equation}
\ket{\vec{\sigma}}=\ket{\sigma_{1}}\otimes \ket{\sigma_{2}}\otimes \cdots \otimes \ket{\sigma_{N}}, \label{configuration}
\end{equation}
and $\ket{\sigma_{j}}$ the eigenstate of $\sigma_{j}^{z}$. The dimension of the Hilbert space is $M=2^N$.

\subsection{Exact diagonalization}
To achieve exact diagonalization, one has to write the Hamiltonian matrix in the basis (\ref{configuration}) and hence sort these vectors. A way to do that is to use the binary code of a configuration: when a spin is \textit{up} (resp. \textit{down}) we replace it by a $1$ (resp. $-1$). For $N=2$ one has for instance
\begin{equation}
\left\{
\begin{array}{rl}
  0&=\ket{0}\equiv\ket{\downarrow \downarrow}=  \ket{-1-1}, \\
1&=\ket{1}\equiv \ket{\uparrow \downarrow}=\ket{\hspace{4.5mm} 1 -1},\\
  2&=\ket{2}\equiv \ket{\downarrow \uparrow }= \ket{-1 \hspace{4mm} 1},\\
  3&=\ket{3}\equiv \ket{\uparrow \uparrow}=\ket{\hspace{4.5mm}1 \hspace{4mm} 1}.
\end{array}
\right. \label{exemple_base}
\end{equation}
Consequently, there is a bijection between all integer numbers between $0$ and $2^N-1$ and  configurations of the spins chain. The element of the Hamiltonian can then be written:
\begin{equation}
H_{ij}\equiv \bra{i}\mathcal{H}\ket{j},\quad i,j=0,1,2,\cdots,2^{N}-1.
\end{equation}
The terms in  \eqref{general_H} which contain $\sigma_n^x\sigma_{n+1}^x$ and $\sigma_n^{y}\sigma_{n+1}^y$  flip two adjacent spins at site $n$ and $n+1$. The term $\sigma_n^{x}$ flips spin at site $n$ only. These terms give rise to off-diagonal elements of the Hamiltonian matrix. The terms 
$\sigma_n^z\sigma_{n+1}^z$ and $\sigma_n^z$ give contribution to its diagonal part. 

Once the Hamiltonian matrix has been defined, one can use standard diagonalization library and compute eigenvalues and eigenvectors. Typically, these algorithm take a time proportional to $M^3$ so that  one can only attain easily spin chains up to $13$ spins.  We shall see later that this is sufficient to compute fractal dimensions  with relatively good precision. However, it is necessary to adopt an iterative (and so approximate) method of diagonalization in order to reach spin's chain with large number of spins. 

\subsection{Lanczos technique}

The Hamiltonian matrix is a very sparse matrix: it contains in each row and each column $K \sim N \ll M=2^N$ non-zero matrix elements . The Lanczos algorithm is an iterative algorithm based on power methods \cite{Lanczos}. It permits to find lowest (or largest) eigenvalues and corresponding eigenvectors of a square matrix and is particularly useful for finding decompositions of very large sparse matrices. 

Basically, the Lanczos algorithm construct a special basis, the so called Krylov space, where the Hamiltonian has a tridiagonal representation. The algorithm works as follow (see e.g. \cite{dagotto}): select an arbitrary vector $\ket{\phi_{0}}$ in the Hilbert space of the model being studied. Generally, this vector is chosen randomly to ensure that the overlap with the actual ground state is non-zero. If some symmetries of the ground state are known, then it is convenient to initiate the iterations with a state already belonging to the subspace having those quantum numbers. We then create a new vector by applying $\mathcal{H}$ to $\ket{\phi_{0}}$ and subtracting it projection over $\ket{\phi_{0}}$:

\begin{equation}
\ket{\phi_{1}}=\mathcal{H}\ket{\phi_{0}}-\frac{\bra{\phi_{0}}\mathcal{H}\ket{\phi_{0}}}{\psh{\phi_{0}}{\phi_{0}}} \ket{\phi_{0}},
\end{equation}
which satisfies $\psh{\phi_{0}}{\phi_{1}}=0$. Doing the same thing with $\ket{\phi_{1}}$, we construct a new state which is orthogonal to the two previous one:
\begin{equation}
\ket{\phi_{2}}= \mathcal{H}\ket{\phi_{1}}-\frac{\bra{\phi_{1}}\mathcal{H}\ket{\phi_{1}}}{\psh{\phi_{1}}{\phi_{1}}} \ket{\phi_{1}}-\frac{\psh{\phi_{1}}{\phi_{1}}}{\psh{\phi_{0}}{\phi_{0}}}\ket{\phi_{0}}.
\end{equation}

Generalizing to the order $n$, we have:
\begin{equation}
\ket{\phi_{n+1}}=\mathcal{H}\ket{\phi_{n}}-a_{n}\ket{\phi_{n}}-b_{n}^{2}\ket{\phi_{n-1}}, 
\end{equation}
where $n=0,1,2,\dots$ and the coefficients are given by:
\begin{equation}
a_{n}=\frac{\bra{\phi_{n}}\mathcal{H}\ket{\phi_{n}}}{\psh{\phi_{n}}{\phi_{n}}}, \hspace*{0.1cm} b_{n}^{2}=\frac{\psh{\phi_{n}}{\phi_{n}}}{\psh{\phi_{n-1}}{\phi_{n-1}}}
\end{equation}
with the condition $b_{0}=0$. It is easy to check that the vector created at step $n$ is orthogonal to the $n-1$ previous one. After $n$ iterations the Hamiltonian matrix  has the following tridiagonal form in the basis $\lbrace \ket{\phi_{k}}\rbrace_{0 \leq k \leq n}$

\[
\hspace*{0.4cm}
T_{n}= \begin{pmatrix}
 a_{0} &  b_{1} &  0 & \cdots & 0 \\
 b_{1} & a_{1} & b_{2} & \ddots & \vdots \\
 0  &  b_{2} &  \ddots & \ddots &  0\\
 \vdots & \ddots &  \ddots & a_{n-1} & b_{n}\\
 0 & \cdots & 0 & b_{n} & a_{n}
  \end{pmatrix}.
\]

Once in this form the matrix can be diagonalized easily using standard library subroutines. A number of iterations equal to the size of the Hilbert space is necessary to diagonalize exactly the model being studied.
However, a number $n \sim 100 \ll M$ iterations is sufficient in practice to have good enough accuracy for lowest energy states. The ground state wave function in (\ref{psi_expansion}) is expressed in the $\lbrace \ket{\phi_{k}}\rbrace_{0 \leq k \leq n}$ basis as $\Psi=\sum_{m}c_{m}\ket{\phi_{m}}$ and the coefficient $c_{m}$ are obtained during the diagonalization of $T_{n}$. By this method in desktop computers one can easily find lowest states for spin chains up to $20$ spins.

\section{Quantum Ising model}\label{QIM}

The quantum Ising model in transverse field \cite{pfeuty} is a well studied model of quantum phase transitions \cite{sachdev}. It is defined by the Hamiltonian \eqref{general_H} with $\Delta=\alpha=0$ and $\gamma=1$:
\begin{equation}
\mathcal{H}_{\mathrm{QIM}}=-\sum_{n=1}^N\left(\sigma_{n}^{x}\sigma_{n+1}^{x}+
\lambda\sigma_{n}^{z}\right). 
\label{Ising_Hamiltonian}
\end{equation}
We consider the case of ferromagnetic Ising model with periodic boundary conditions, $\sigma_{N+1}=\sigma_1$.
\subsection{Analytical results} 

\begin{figure}[!ht]
\begin{center}
\includegraphics[width=.3\linewidth, clip]{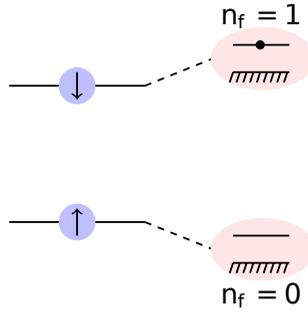}
\end{center} 
\caption{(Online version in colour.) Key idea of Jordan-Wigner transformation: an up-spin $\ket{\uparrow}$ (resp. down spin $\ket{\downarrow}$) is equivalent to absence $n_{f}=0$  (resp. the presence $n_{f}=1$) of a fermion.}
\label{jordan_wigner}
\end{figure}

The spectrum of this model can be found analytically  by the Jordan-Wigner transformation \cite{lieb} which maps the spin problem into a free fermions problem (see Fig.~\ref{jordan_wigner}).

The eigen-energies of the QIM are given by the fermionic filling of one-particle levels 
\begin{equation}
E=\sum_k e_k\left (n_k-\tfrac{1}{2}\right )
\label{excitation_spectrum_ising}
\end{equation}
where $n_k=0,1$ is the number of fermions with energy $e_k$ given by
\begin{equation}
e_k =2\sqrt{1-2\lambda\cos k+\lambda^{2}} \ ,
\label{excitation_energy_ising}
\end{equation}
with $k$ taking $N$ values depending on the parity of the excitation
\begin{equation}
P=(-1)^{n_{\mathrm{down}}}
\label{parity}
\end{equation}
where $n_{\mathrm{down}}$ is the number of spins down. 

One has \cite{lieb}
\begin{equation}
k=\frac{2\pi}{N}\left \{ \begin{array}{ll}l,& P=1\\l+\tfrac{1}{2},&P=-1\end{array}\right .,\qquad l=0,1,\ldots,N-1\ .
\label{k}
\end{equation}
The ground state energy corresponds to all $n_k=0$. 

The calculation of the ground state eigenfunction is more involved due to the necessity to perform the Bogoliubov transformation \cite{lieb}, \cite{pasquier_1,pasquier_2}. 
For each sequence of $\sigma$'s  coefficients $\Psi_{\sigma}$ in  \eqref{psi_expansion} for the GS are given by the determinant of  $N \times N$ matrix
\begin{equation}
|\Psi_{\vec{\sigma}}|^{2}=\det \left [ \frac{1}{2}\Big (\delta_{mn}-\sigma_m O_{mn}\Big )\right ]_{m,n=1,\ldots, N} 
\label{configuration_probability}
\end{equation}
with the following  orthogonal matrix $O_{mn}$  \cite{lieb}, \cite{pasquier_1}:
\begin{equation}
O_{mn}=\frac{1}{N} \sum_{k}\cos(k(m-n)+2\theta_{k})\ ,
\label{matix_proba}
\end{equation}
where the angle $\theta_{k}$ ($\theta_{-k}=\theta_{k}$) is
\begin{equation}
\cos 2\theta_{k}=\frac{\cos k-\lambda}{\sqrt{1-2\lambda\cos k+\lambda^{2}}}\ . 
\label{angle_Ising_PBC}
\end{equation}
For the GS in  the QIM   $k=2\pi(l+1/2)/N$ and the summation over $k$ indicates the sum over all $l=0,1,\ldots, N-1$.

\begin{figure}[!ht]
\begin{minipage}{.49\linewidth}
\begin{center}
\includegraphics[angle=-90, width=.99\linewidth, clip]{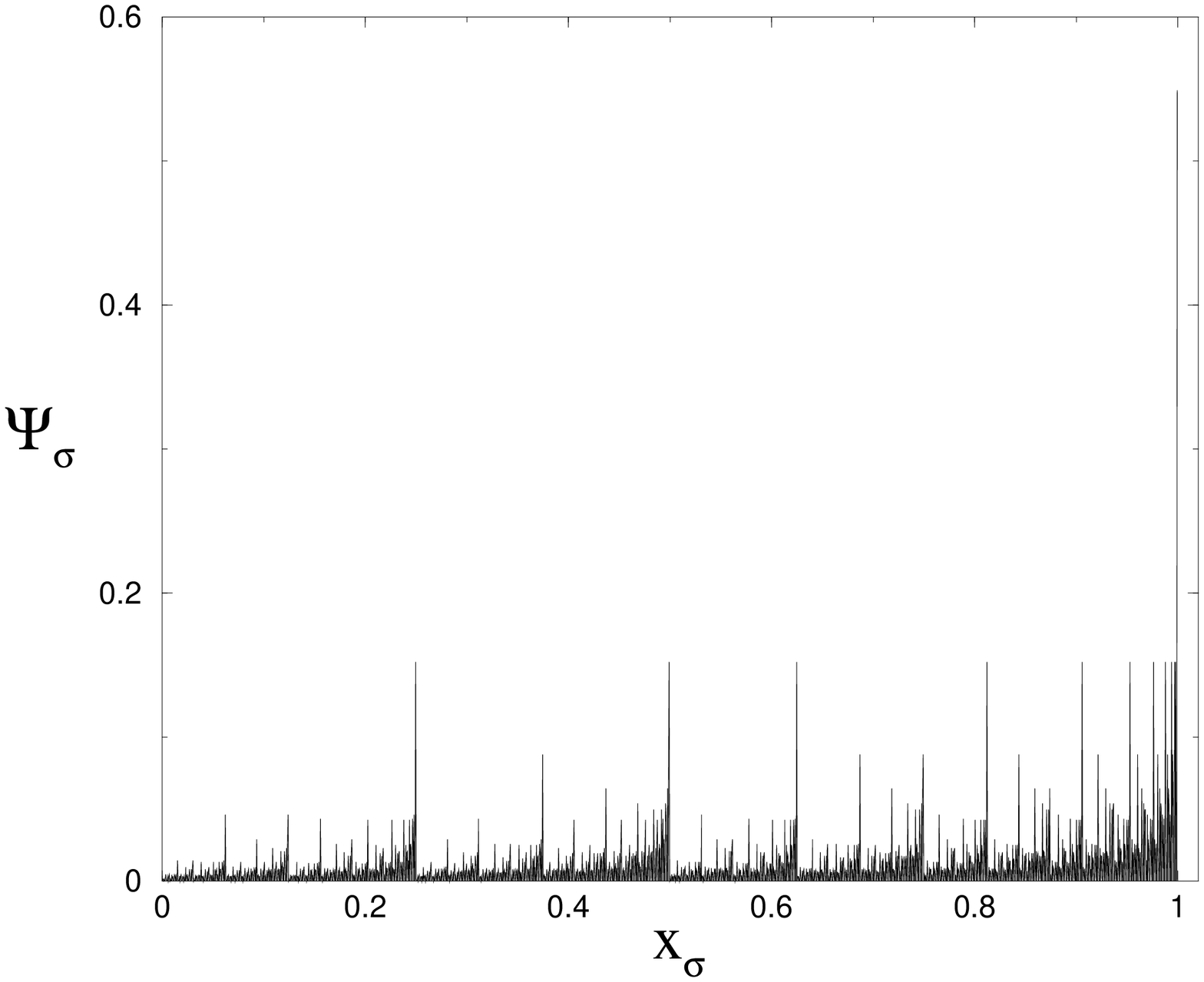}\\ 
(a) \end{center}
\end{minipage}
\begin{minipage}{.49\linewidth}
\begin{center}
\includegraphics[angle=-90, width=.99\linewidth, clip]{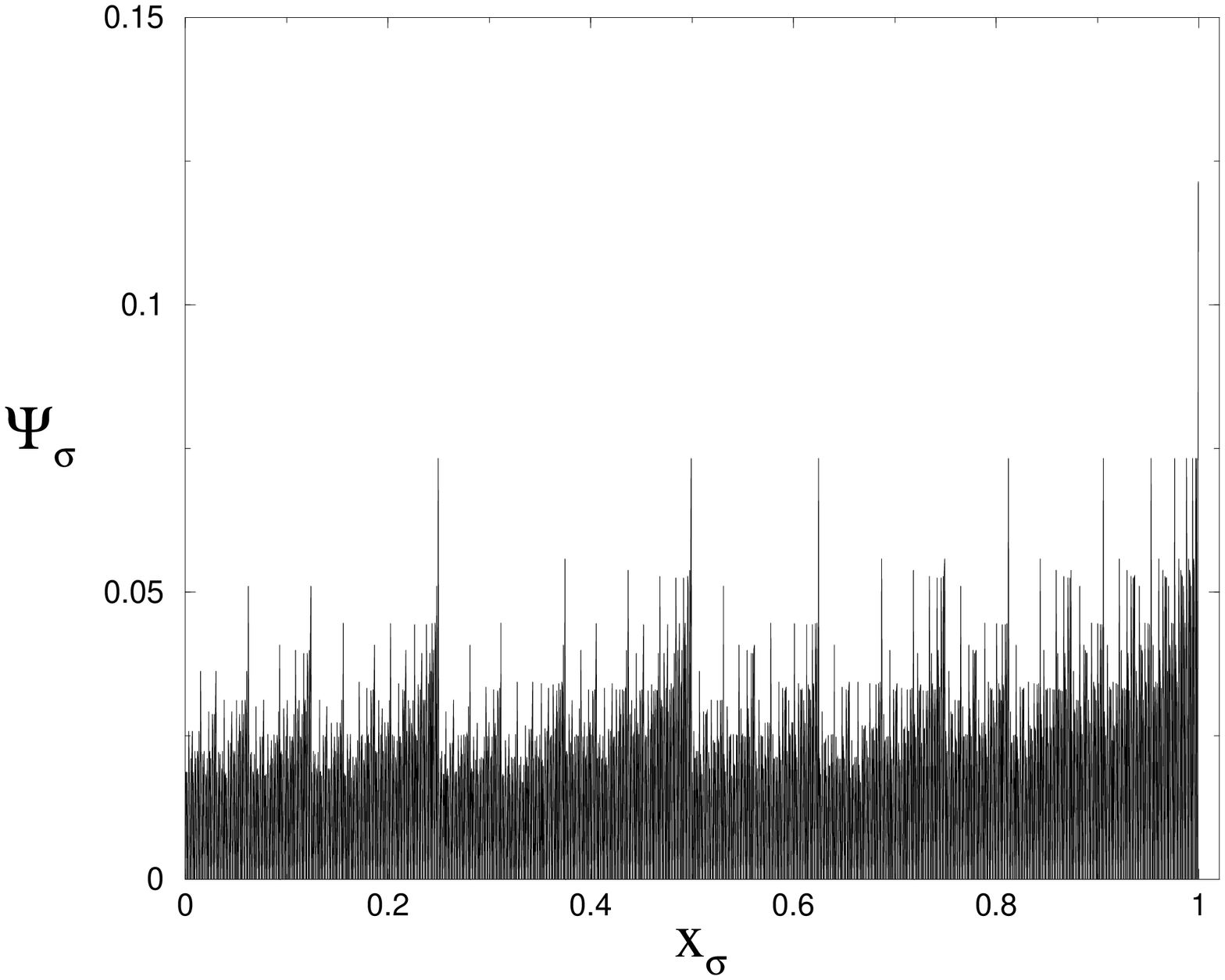} \\
(b) \end{center}
\end{minipage}
\caption{(a) Coefficients of the GS wave function for the QIM with $\lambda=1$ and $N=11$ versus the binary code \eqref{binary_code}. (b) The same but for the XY model  with $\lambda=0.4$, $\gamma=1.4$, and $N=12$. }
\label{wave_function_Ising}
\end{figure}

These expressions are useful for numerical calculations of ground state properties  as well as the asymptotics of fractal dimensions when $q\to\pm \infty$ \cite{pasquier_1,pasquier_2}. For illustration in Fig.~\ref{wave_function_Ising} a) coefficients  
of the GS function for the critical QIM with $\lambda=1$ are presented. 

To find $D_{\infty}$ it is necessary to know the largest coefficient in the expansion \eqref{psi_expansion}. From physical considerations it is clear that it  corresponds to the pure  ferromagnetic configuration where all the spins are up. Eq.~\eqref{configuration_probability} when all $\sigma=1$ gives 
\begin{equation}
|\Psi_{\uparrow\uparrow\cdots\uparrow}|^2=\prod_{k}\sin^2 \theta_k \ . 
\label{proba_ferro}
\end{equation}
When $N\to\infty$ one obtains \cite{pasquier_2}
\begin{equation}
D_{\infty}(\lambda)=\frac{1}{2}-\frac{1}{2\pi \ln 2}\int_0^{\pi}\ln \Big [1+ \frac{\lambda-\cos u}{\sqrt{1-2\lambda \cos u +\lambda^2 }} \Big ]\mathrm{d}u \ . \label{D_p}
\end{equation}
Using similar arguments one concludes that $D_{-\infty}$ is related with the minimal coefficient in \eqref{psi_expansion}. As we consider the ferromagnetic Ising model it is plain that the minimum is attained for a configuration with the maximum number of spins down. If $N$ is even then the configuration \eqref{configuration_probability} when all $\sigma=-1$ is allowed. For odd $N$ there exist $N$ configurations with $N-1$ spin down and one spin up. In all cases when $N\to\infty$ one concludes \cite{pasquier_2} that 
\begin{equation}
D_{-\infty}(\lambda)=\frac{1}{2}-\frac{1}{2\pi \ln 2}\int_0^{\pi}\ln \Big [1-\frac{\lambda-\cos u}{\sqrt{1-2\lambda \cos u +\lambda^2}} \Big ]\mathrm{d}u \ . 
\label{D_m}
\end{equation}
Using the Kramers-Wannier duality in \cite{pasquier_2} it was shown that for the QIM it is possible to calculate also $D_{1/2}$. Generalizing slightly the arguments of this work we get
\begin{equation}
D_{1/2}(\lambda)=1-D_{\infty}\Big (\frac{1}{\lambda}\Big )\ . 
\label{D_1/2}
\end{equation}
These formulas prove that fractal dimensions of quantum Ising model do exist and are non-trivial. 

The above formulae are qualitatively the same for non-critical and critical (that is $\lambda=1$) QIM but their sum  
\begin{equation}
D_{-\infty}(\lambda)+D_{\infty}(\lambda)=\left \{ \begin{array}{cc}2, &|\lambda|<1\\2+\frac{\ln |\lambda|}{\ln 2}, &|\lambda|>1\end{array}\right .
\label{D_minus}
\end{equation}
has singularity at $\lambda=1$ which means that fractal dimensions are also sensitive to quantum criticality.
\subsection{Numerics}

As was mentioned above the simplest method to find fractal dimensions is the direct numerical calculation of the GS wave function for different number of spins and a subsequent extrapolation of the R\'enyi entropy for large $M$. In the definition \eqref{fractal} it is implicitly assumed that $q$ is positive. For many problems (but not for all) fractal dimensions  can also   be calculated for negative $q$  \cite{mandelbrot_2}, \cite{riedi}. When certain coefficients in \eqref{psi_expansion} are zero due to an exact  symmetry (as in QIM), they are not included in  the calculation of the R\'enyi entropy \eqref{renyi_entropy} for $q\leq 0$.  The curves $S_{R}(q,M)$ as a functions of $z=\ln M=N\ln 2$ were fitted to the simplest functional form
\begin{equation}
 f(z)=a_{0}+a_{1}z+\frac{a_{2}}{z}+\frac{a_{3}}{z^2}. 
 \label{fit_f}
\end{equation} 
The coefficients $a_{i}$ are, of course, functions of $q$. In practically all models, these curves $\ln P_q$ (cf. \eqref{tau_q}) as function of $N$ are almost straight lines and the slope $a_{1}$ gives fractal dimensions with a reasonable precision. To estimate the precision of the fit we calculate the sum
\begin{equation}
\chi^2(q)=\sum_{M}|\ln P_q(M)-f(N\ln 2)|^2
\label{chi_2}
\end{equation} 
where the summation is performed over all available values of matrix dimensions $M$ and $P_q(M)$ is defined in \eqref{tau_q}.

\begin{figure}[!ht]
\begin{minipage}{.49\linewidth}
\begin{center}
\includegraphics[width=.99\linewidth, clip]{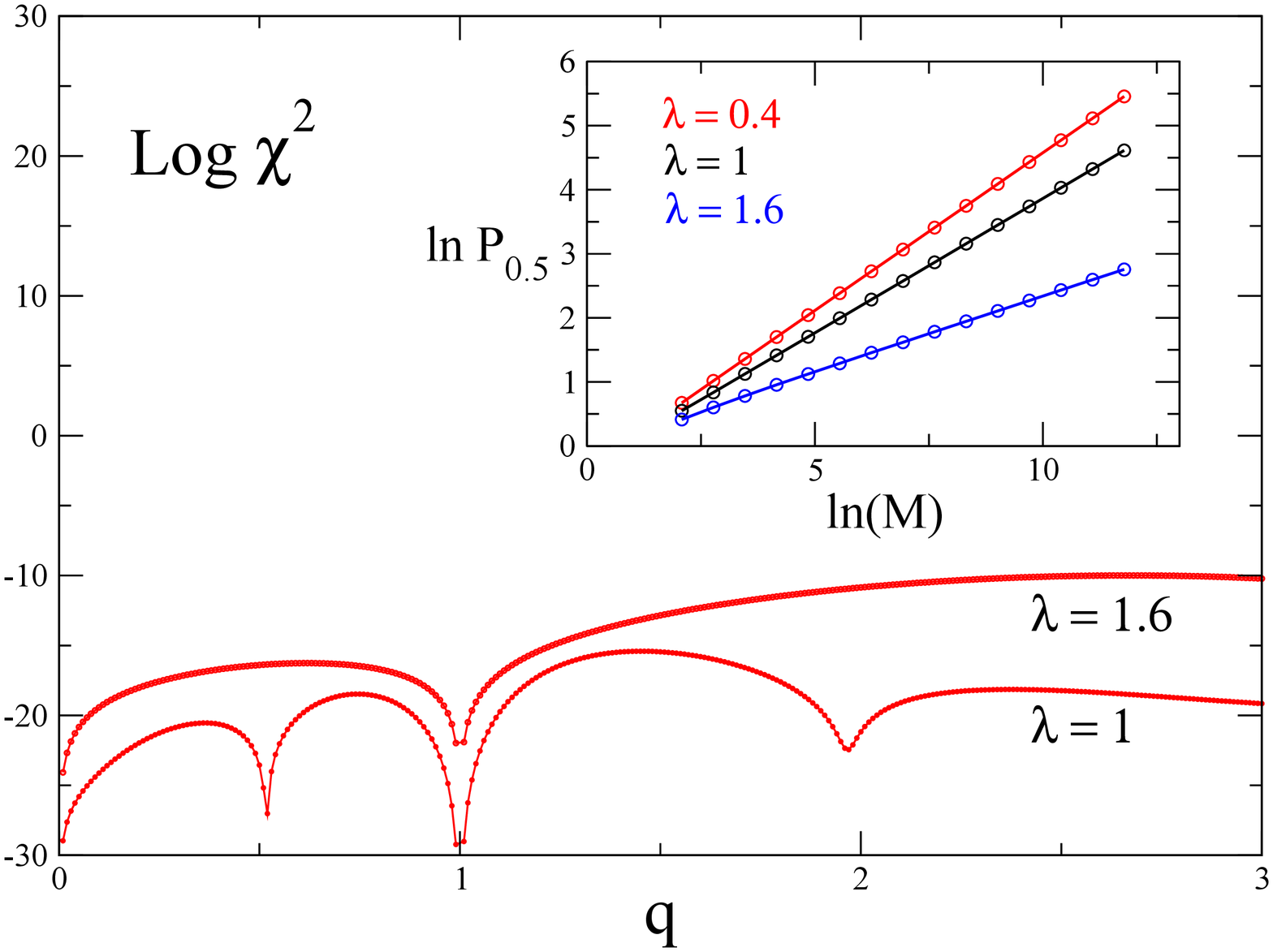}\\
 (a) \end{center}
\end{minipage}
\begin{minipage}{.49\linewidth}
\begin{center}
\includegraphics[width=.99\linewidth ,clip]{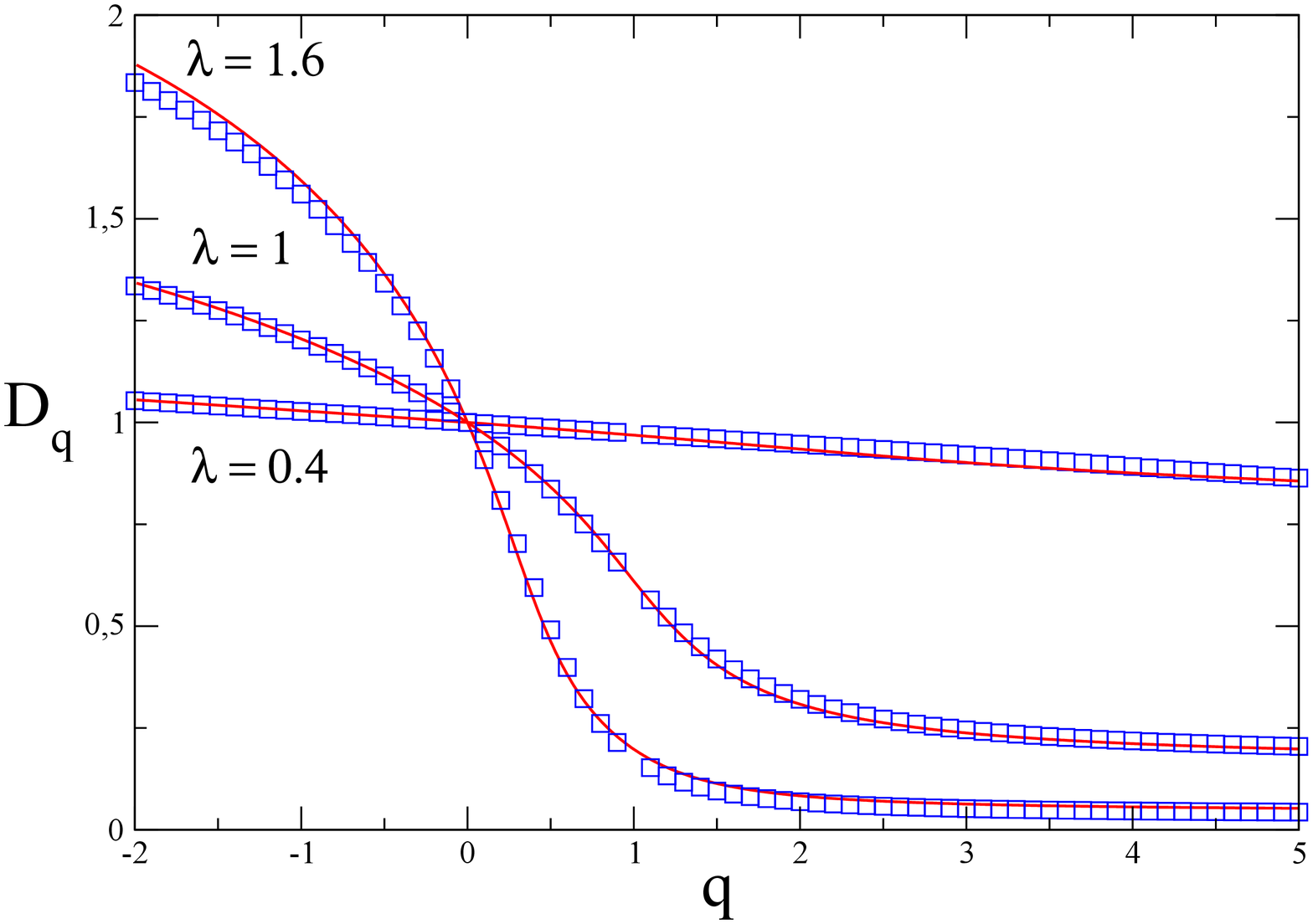} \\
(b) \end{center}
\end{minipage}
\caption{(Online version in colour.) (a) Plot of the logarithm of the chi-square \eqref{chi_2} as a function of  $q$ for  fit  \eqref{fit_f} for the Ising model in critical field $\lambda=1$ and non-critical field $\lambda=1.6$. Inset: $\ln P_{1/2}$ as function of $\ln M=N\ln 2$ for $\lambda=0.4,\;1,\;1.6$. (b) Numerical comparison of the fractal dimensions for the critical quantum Ising model obtained by Lanczos technique with $N=3-18$ (red lines) and exact diagonalization with $N=3-11$ (blue squares }
\label{error_Ising}
\end{figure}

For the Ising model and all the other models, we perform Lanczos algorithm with $n=150$ iterations. As in the QIM  parity is conserved,  the iteration of the Lanczos method starts with the ferromagnetic configuration with all spins up.
The quality of the  fit is illustrated in  Fig.~\ref{error_Ising}~a). In Fig.~\ref{error_Ising}~b) the comparison between the direct diagonalization and the Lanczos method is presented. Though exact diagonalization was performed for relatively small number of spins, the results agree well with Lanczos results.

As an example of precision of numerical calculations we compare the exact values of $D_{1/2}$  with our numerics. One has from \eqref{D_1/2}
\begin{equation}
D_{1/2}(1)=\frac{2K}{\pi \ln 2}\approx 0.841267,\; D_{1/2}(0.4)\approx 0.984871,\; D_{1/2}(1.6)\approx 0.467946\ .
\end{equation}
Here $K$ is the Catalan constant. 

Numerical fits give
\begin{equation}
D_{1/2}(1)\approx 0.841283,\; D_{1/2}(0.4)\approx 0.984789, \; D_{1/2}(1.6)\approx 0.462985\ .
\end{equation}
We see that the accuracy of the calculations at this particular point is of the order $10^{-3}-10^{-4}$ which is enough for all practical purposes. More numerical results are presented in \cite{ab}.


\section{XY model}\label{XY}

The Hamiltonian of this model differs from the QIM by the anisotropy $\gamma$ 
\begin{equation}
H_{XY}=-\sum_{n}\left( \frac{1+\gamma}{2}\sigma_{n}^{x}\sigma_{n+1}^{x}+\frac{1-\gamma}{2}\sigma_{n}^{y}\sigma_{n+1}^{y}+\lambda\sigma_{n}^{z}\right)\ . 
\label{XY_hamiltonian}
\end{equation}
Similar to the QIM this model is also integrable by the Jordan-Wigner transformation \cite{lieb}, \cite{pasquier_2} but the structure of its GS is more complicated (see below). The eigen-energies are given by the same expression \eqref{excitation_spectrum_ising} but with 
\begin{equation}
e(k)=2\sqrt{(\cos k-\lambda)^2+\gamma^2\sin^2 k}
\end{equation}
where $k$ is as in \eqref{k}.
 
As for the Ising model the GS wave function coefficients are calculated from \eqref{configuration_probability} and \eqref{matix_proba} but with $\theta_k$ determined by the expression  
\begin{equation}
\cos 2\theta_{k}=\frac{\cos k-\lambda}{\sqrt{(\cos k-\lambda)^2+\gamma^2\sin^2 k}}. 
\label{angle_XY}
\end{equation}
The XY model reduces to the Ising model for $\gamma=1$. In Fig.~\ref{wave_function_Ising} b) an example of the GS wave function for the XY model is presented.


\subsection{Ground state energy degeneracy}

Due to  anisotropy $\gamma$ the parity of the GS is not fixed a-priori. Depending on parameters $\lambda$ and $\gamma$ the lowest energy state  may have either odd or even parity. Correspondingly, the GS energy is given by the same formula as \eqref{excitation_spectrum_ising} (with $n_k=0$)
\begin{equation}
E=-\tfrac{1}{2}\sum_k e_k 
\end{equation}
but either with $k=2 \pi l/N$ or $k=\pi (2l+1)/N$ which we called even and odd momenta. 

The energy difference between these two states is
\begin{equation}
E_{\mathrm{even}}-E_{\mathrm{odd}}=-\tfrac{1}{2}\sum_{l=0}^{N-1}\left [e\Big (\frac{2\pi l}{N}\Big )-e\Big (\frac{\pi (2l+1)}{N}\Big )\right ]\ .
\label{difference}
\end{equation}
If this difference is positive, $E_{\mathrm{GS}}=E_{\mathrm{odd}}$, if it is negative, $E_{\mathrm{GS}}=E_{\mathrm{even}}$.   
To calculate the sum in \eqref{difference}  when $N\to\infty$ one can proceed as follows. 
  
\begin{figure}
\begin{center}
\includegraphics[width=.3\linewidth]{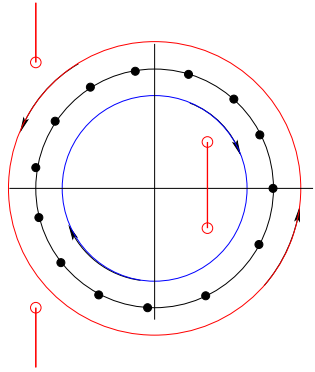} 
\end{center}
\caption{(Online version in colour.) Contour of integration in the calculation of the ground state for the XY model. Black filled circles indicate poles on the unit circle. Red open circles are square root singularities of function $f(z)$ in the case  $\lambda^2+\gamma^2<1$}
\label{contour_fig}
\end{figure}

Denote $z=\mathrm{e}^{\mathrm{i}k}$. For even modes $z_l$ are roots of $z^N-1=0$, for odd modes $z_l$ are roots of $z^N+1=0$. It is easy to check that
\begin{equation}
E_{\mathrm{even}}-E_{\mathrm{odd}}=\frac{N}{\pi \mathrm{i}}\oint_C  \frac{z^{N-1}}{1-z^{2N}}·\, f(z )\, \mathrm{d}z, \qquad f(z)=\frac{1}{2z}\sqrt{P(z)}
\end{equation}
with $P(z)=(z^2 -2\lambda z+1)^2-\gamma^2(z^2-1)^2$  and contour $C$ encircles all poles on the unit circle  as in Fig.~\ref{contour_fig}.

Function $f(z)$ has 4 square root singularities when $P(z)=0$: 
\begin{eqnarray}
z_1&=&\frac{\lambda+\sqrt{\lambda^2+\gamma^2-1}}{1+\gamma}\, ,\qquad z_2=\frac{\lambda-\sqrt{\lambda^2+\gamma^2-1}}{1+\gamma}\, ,\\
z_3&=&\frac{\lambda+\sqrt{\lambda^2+\gamma^2-1}}{1-\gamma}=\frac{1}{z_2}\, ,\qquad z_4=\frac{\lambda-\sqrt{\lambda^2+\gamma^2-1}}{1-\gamma}=\frac{1}{z_1}\, .
\nonumber
\end{eqnarray}
If $\lambda^2+\gamma^2<1$ these roots are complex, when $\lambda^2+\gamma^2\geq 1$ they are real (cf. Fig.~\ref{contour_fig}). Contour $C$ can be deformed to encircle the cuts of $f(z)$. 

When $N\to\infty$  one can use $(z_j+v)^N\approx z_j^N\mathrm{e}^{Nv/z_j}$
and $P(z)$ close to $z=z_1$ can be approximated as
$P(z)=P^{\prime}(z_1)(z-z_1)$
where
\begin{equation}
P^{\prime}(z_1)=(1-\gamma^2)(z_1-z_2)(z_1-z_3)(z_1-z_4)
\end{equation}
with the similar formulae for other zeros. 

Performing straightforward calculations one gets
\begin{equation}
E_{\mathrm{even}}-E_{\mathrm{odd}}=-\mathrm{Re} \left [r^{N}(\lambda,\gamma) \sqrt{\frac{32(1-\lambda^2-\gamma^2)}{\pi N(1-\gamma^2)}\Big (\sqrt{1-\lambda^2-\gamma^2}+\mathrm{i}\gamma\lambda }\Big )\, \right ]
\label{difference_pm}
\end{equation}
where 
\begin{equation}
r(\lambda,\gamma)=\frac{\lambda+\mathrm{i}\sqrt{1-\lambda^2-\gamma^2}}{1+\gamma}\ .
\end{equation}
When $\lambda^2+\gamma^2<1$  the difference \eqref{difference_pm} oscillates (it has $N$ zeros) and  is exponentially small,   $E_{\mathrm{even}}-E_{\mathrm{odd}}\sim |r|^N$ where
\begin{equation}
|r|=\sqrt{\frac{1-\gamma}{1+\gamma}}.
\end{equation}
When $\lambda^2+\gamma^2<1$  this difference is positive and $E_{\mathrm{GS}}=E_{\mathrm{odd}}$ as for the QIM. 
 
To avoid the difficulty of finding the correct ground state, we choose parameters $\lambda$ and $\gamma$ outside or in the unit circle
\begin{equation}
\lambda^2+\gamma^2 \geq  1
\end{equation}
and performs  calculations of fractal dimensions only in such case.

\subsection{Factorizing field}

When $\lambda^2+\gamma^2=1$ or $\lambda=\lambda_f$ where
\begin{equation}
\lambda_f=\sqrt{1-\gamma^2}\, ,
\end{equation}
the XY model has two degenerate GS with different parity. 

It is known \cite{kurmann} that at that field the XY model has two exact factorized GS wave functions
\begin{equation}
\Psi(\theta)=\prod_{n=1}^N(\cos \theta \ket{\uparrow}_{n} + \sin \theta \ket{\downarrow}_{n}),
\label{factorized}
\end{equation}
with 
\begin{equation}
\cos^2 2\theta=\frac{1-\gamma}{1+\gamma}\ .
\label{factor_angle}
\end{equation}
This ground state is doubly degenerated as the both $\theta$ and $-\theta$ obey Eq.~\eqref{factor_angle}.

For completeness we sketch here the proof of this result.  To find the factorizing field in the XY model it is sufficient to consider the one term in the Hamiltonian \eqref{XY_hamiltonian}  (so that $\sum_n \mathcal{H}_n=\mathcal{H}$)
\begin{equation}
\mathcal{H}_n=- \tfrac{1}{2} (1+\gamma)\sigma_n^{x}\sigma_{n+1}^x-\tfrac{1}{2}(1-\gamma)\sigma_n^{y}\sigma_{n+1}^y  
-\tfrac{1}{2}\lambda  (\sigma_n^z+\sigma_{n+1}^z )
\end{equation}
and try to fulfil the eigenvalue condition 
\begin{equation}
\mathcal{H}_n \Psi_n=e \Psi_n
\label{factorization}
\end{equation}
with 
\begin{equation}
\Psi_n=\big [\cos\theta \ket{\uparrow}_n +\sin \theta \ket{\downarrow}_n \big] \big [\cos \theta \ket{\uparrow}_{n+1}+\sin \theta \ket{\downarrow}_{n+1} \big]\ . 
\end{equation} 
Direct calculations give that Eq.~\eqref{factorization} will be fulfilled iff the parameters obey the equations   
\begin{equation}
e\,  \cos^2\theta =  -\gamma \sin^2\theta -\lambda \cos^2\theta ,\;
e\,  = -1,\;
e\,  \sin^2\theta = -\gamma \cos^2\theta +\lambda  \sin^2\theta .
\end{equation}
These equations are easily solved 
\begin{equation}
\lambda_f=\sqrt{1-\gamma^2} 
\end{equation}
and they give Eq.~\eqref{factor_angle}.

As $e=-1$ the full GS energy for such factorized  state is
\begin{equation}
E_f=-N .
\end{equation}
In the factorizing state the coefficients of the expansion \eqref{psi_expansion}  are
\begin{equation}
\Psi_{\sigma}=\cos^n \theta \sin^{N-n}\theta 
\end{equation}
where $n$ is the number of spins up in the state $\sigma$. Comparing it with  \eqref{proba_binomial} one concludes that it corresponds exactly to the binomial measure discussed in Section~\ref{sec_binomial}.  Therefore, fractal dimensions in this case are (cf. \eqref{binomial_dimension})
\begin{equation}
D_q=-\frac{\ln( \cos^{2q}\theta+\sin^{2q}\theta)}{(q-1)\ln 2}\ .
\label{xy_exact}
\end{equation}
To remove the doubly degeneration of the GS in the factorizing field it is convenient to form two states with different parities \eqref{parity}
\begin{equation}
\Psi^{(\pm)}=\tfrac{1}{\sqrt{2N_{\pm}}}[\Psi(\theta)\pm \Psi(-\theta)]\ .
\end{equation} 
where $N_{\pm}$ are  normalization constants
\begin{equation}
N_{\pm}=1\pm \cos^N2\theta \ .
\end{equation} 
For such states with fixed parity
\begin{equation}
P_q\equiv \sum_{\sigma} |\Psi_{\sigma}^{(\pm)} |^{2q}=\frac{2^{q-1}}{N_{\pm}^q}\Big [(\cos^{2q}\theta+\sin^{2q}\theta)^N \pm (\cos^{2q}\theta-\sin^{2q}\theta)^N \Big ]
\label{exact_p_q}
\end{equation}
and the fractal dimensions are the same as in \eqref{xy_exact}.

\begin{figure}[t!]
\begin{minipage}{.49\linewidth}
\begin{center}
\includegraphics[width=.99\linewidth,clip]{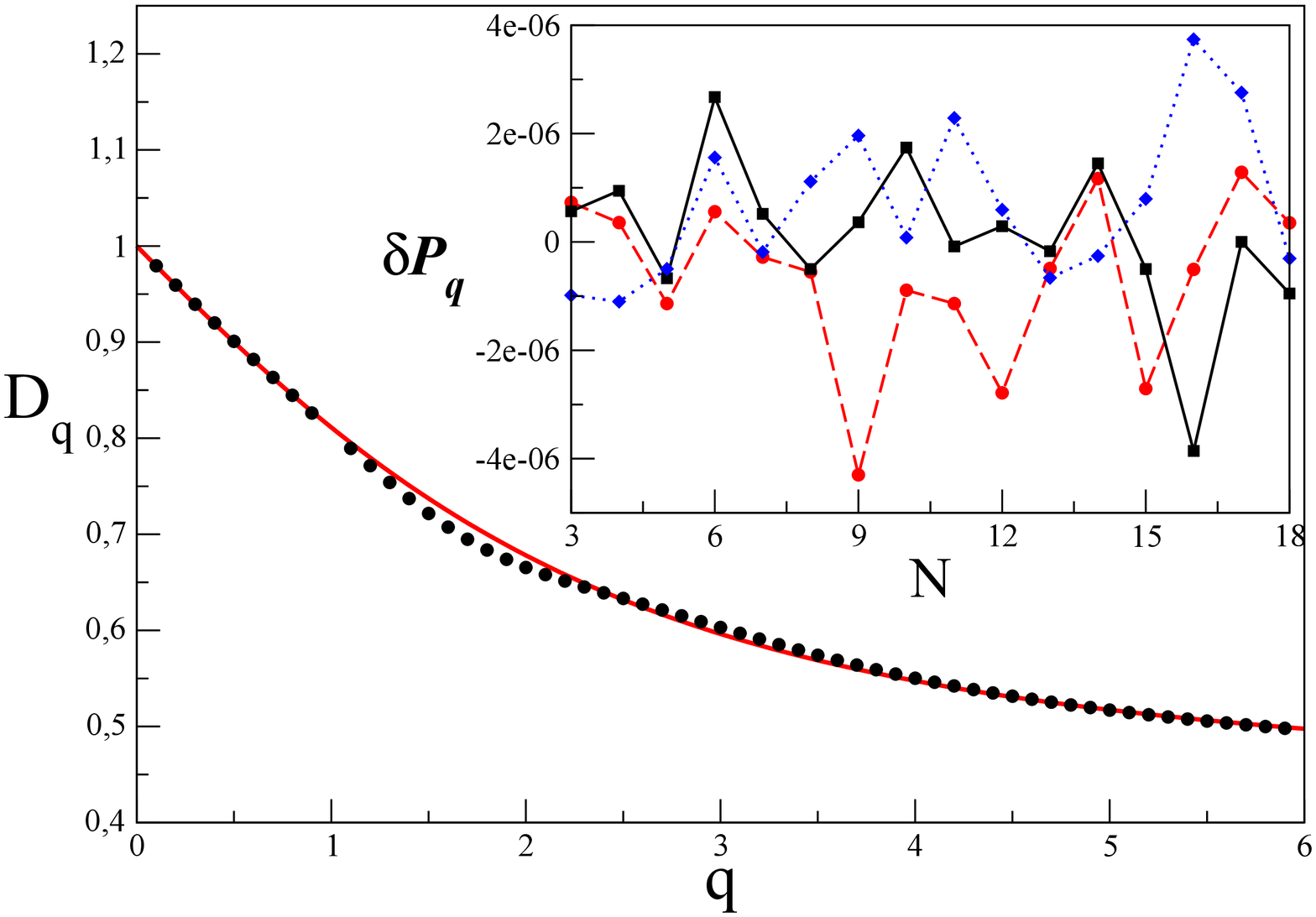} \\
(a) \end{center}
\end{minipage}
\begin{minipage}{.49\linewidth}
\begin{center}
\includegraphics[width=.99\linewidth,clip]{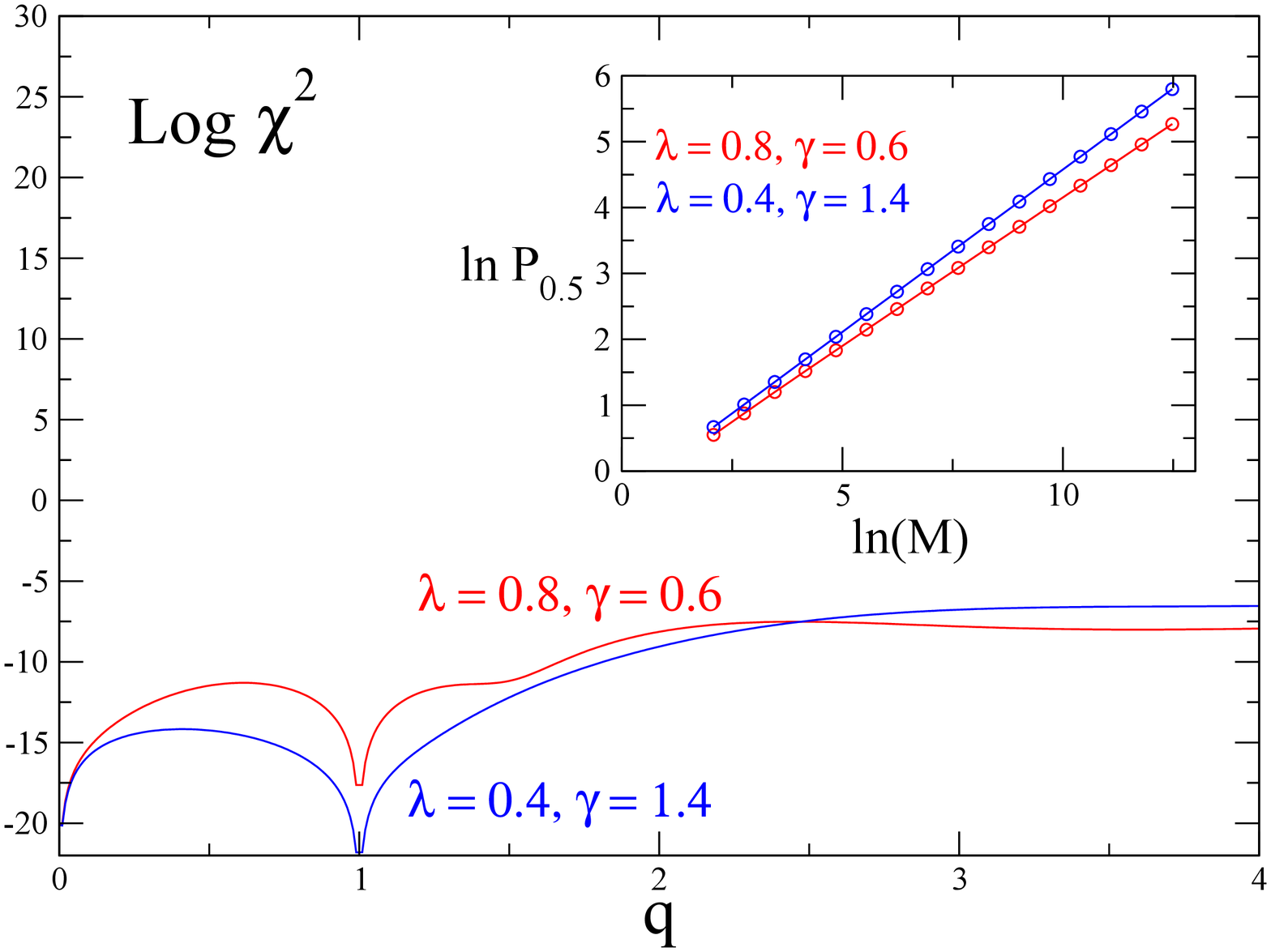} \\
(b) \end{center}
\end{minipage}
\caption{(Online version in colour.) (a) Fractal dimensions of the GS for the XY model in the factorising field $\lambda=0.8$ and $\gamma=0.6$ obtained by  Lanczos method  for $N=5-18$ are indicated by black circles. The red solid line corresponds to the exact Eq.~\eqref{xy_exact} with  $\theta=\pi/6$. Inset: Relative errors of moments of GS wave function \eqref{error}  at these parameters: black squares correspond to $q=2$, red circles - to $q=2.5$, and blue rhombuses - to $q=3.5$. (b) Plot of the logarithm of the chi-square \eqref{chi_2} as a function of  $q$ for the XY model in factorising field $\lambda=0.8$ and $\gamma=0.6$, and  generic field $\lambda=0.4$ and $\gamma=1.4$ for the fit \eqref{fit_f}. Inset: $\ln P_{1/2}$ as function of $\ln M=N\ln 2$ for these values of parameters.}
\label{Binomial_XY_dim}
\end{figure}

To illustrate this case, we computed numerically by the Lanczos method moments of GS wave function \eqref{tau_q} for the XY model in the factorising field $\gamma=0.6$ and $\lambda=0.8$ ($\lambda^2+\gamma^2=1$) and compare them with the exact expression \eqref{exact_p_q}. In Inset of Fig.~\ref{Binomial_XY_dim}~a) the relative errors
\begin{equation}
\delta P_q=\frac{(P_q)_{\mathrm{num}}-(P_q)_{\mathrm{exact}}}{(P_q)_{\mathrm{num}}}
\label{error}
\end{equation}
are presented for $q=2,2.5,3.5$. Here $(P_q)_{\mathrm{num}}$ are numerically calculated moments and $(P_q)_{\mathrm{exact}}$ are exact moments  given by \eqref{exact_p_q}. 

The good agreement observed even at large $N$ confirms the precision of numerical calculations. Fractal dimensions at these values of parameters  are calculated using the fit \eqref{fit_f}.  From   Fig.~\ref{Binomial_XY_dim}~b) one can estimate the accuracy of the interpolation of such fit for certain values of parameters in the XY models. Though the fit is quite good (i.e. the fitting points are close to the numerical values), fractal dimensions deviate from the exact ones \eqref{xy_exact}  by  values of the order of $10^{-2}$ as shown in  Fig.~\ref{Binomial_XY_dim}~a). The main reason of such discrepancies is non-uniform convergence of exact formula \eqref{xy_exact} when $N\to\infty$  in different intervals of $q$. As the form of corrections to limiting values for fractal dimensions is unknown, it is difficult, in general, to estimate the precision of calculation of fractal dimensions. From our experience we find that it is reasonable to get the absolute error of the order of $10^{-2}-10^{-3}$ from the data up to $16$ spins though for certain values of $q$ higher accuracy may be obtained.   

\subsection{Limiting values of fractal dimensions}

It seems natural to think that the limiting values, $D_{\infty}$ and $D_{-\infty}$ as in the QIM  should correspond to configurations with, respectively, all spins up and all spins down, and, consequently, are expressed  similar to \eqref{D_p} and \eqref{D_m} as
\begin{equation}
D_{\pm }(\lambda, \gamma)=\frac{1}{2}-\frac{1}{2\pi \ln 2}\int_0^{\pi}\ln \Big [1\pm \frac{\lambda-\cos u}{\sqrt{(\lambda- \cos u)^2 +\gamma^2 \sin^2 u}} \Big ]\mathrm{d}u 
\label{D_limit_XY}
\end{equation}
But it appears that at small $\lambda$ there exits an other distribution of spins which gives the contribution smaller than the one with all spins down.  It corresponds to  the anti-ferromagnetic N\'eel configuration with alternating spins, $\sigma_n=(-1)^n$  whose contribution can be calculated analytically as it is done below. 

For simplicity we consider even $N$. By a simple transformation  from \eqref{configuration_probability} one gets that for the  N\'eel configuration it is necessary to calculate the following determinant
\begin{equation}
|\Psi_{\mathrm{Neel}}|^2=2^{-N} \det( K_{mn} )_{m,n=1,\ldots, N},\qquad  K_{mn}= -(-1)^m \delta_{mn}+ O_{mn}  
\label{Neel}
\end{equation}
where  orthogonal matrix $O_{mn}$ from \eqref{matix_proba} and \eqref{angle_XY} is a  Toeplitz matrix,  $O_{mn}=O_{m-n}$
\begin{equation}
O_r=\frac{1}{N}\sum_k \cos(kr +2\theta_k) =\frac{1}{N} \sum_k \mathrm{e}^{\mathrm{i} kr } \frac{\cos k-\lambda+\gamma \mathrm{i}\sin k}{|\cos k-\lambda+\mathrm{i}\gamma \sin k|}\ .
\end{equation}
Here $k=2\pi (l+1/2)/N$.

When $N\to\infty$ one can change the summation over $k$ into the integration and 
 \begin{equation}
O_{r} =\frac{1}{2\pi} \int_0^{2\pi}\mathrm{d}\phi  \mathrm{e}^{\mathrm{i} r\phi  } g(\phi) ,\;
g(\phi)=\frac{\cos \phi -\lambda +\gamma \mathrm{i}\sin \phi }{|\cos \phi-\lambda+\mathrm{i}\gamma \sin \phi|}\ .
\end{equation}
Matrix $K_{mn}$ in Eq.~\eqref{Neel} is not a Toeplitz matrix but can be written as a  block Toeplitz matrix
\begin{equation}
 K=\begin{pmatrix}
 \Pi_{0} & \Pi_{-1} & \Pi_{-2} &\cdots   \\
  \Pi_{1} & \Pi_{0} & \Pi_{-1} &\cdots  \\
  \Pi_{2} & \Pi_{1} & \Pi_{0} & \cdots \\
  \vdots  & \vdots  & & \ddots  \\
   &  &  &  & \Pi_{0}
 \end{pmatrix}, 
\label{bloc_toeplitz}
\end{equation}
where the  $ \Pi_{k}$ are $2 \times 2$ matrices:
\begin{equation}
 \Pi_{k}=
 \begin{pmatrix}
 	O_{2k} & O_{2k-1}  \\  O_{2k+1} & O_{2k}
 \end{pmatrix}, \; k=\pm 1,\pm 2,\ldots, \; \;  \; \Pi_{0}=
 \begin{pmatrix}
 	-1+O_{0} & O_{-1}  \\ O_{1} & 1+O_{0}
 \end{pmatrix} \, . 
 \label{lespik}
\end{equation}
The asymptotics of the determinant of a block Toeplitz matrices under certain conditions is given by the formula  \cite{widom} (see also \cite{its})
\begin{equation}
\ln \det(K)\to \frac{N}{4\pi}\int_0^{2\pi}\ln\det (\Phi(\theta))\mathrm{d}\theta, \qquad N\to\infty  
\end{equation}
where $\Phi(\theta)$ is the symbol of the block Toeplitz matrix
\begin{equation}
\Phi(\theta)=\sum_{l=-\infty}^{\infty}\mathrm{e}^{\mathrm{i}l\theta}\Pi_{l}=
 \begin{pmatrix}
 	\Phi_{11}(\theta) & \Phi_{12}(\theta) \\  \Phi_{21}(\theta) & \Phi_{22}(\theta)
 \end{pmatrix}\ .
\end{equation}
For matrices $\Pi_k$  given by \eqref{lespik} one gets
\begin{eqnarray}
\Phi_{11}(\theta)&=&-1+\sum_{l=-\infty}^{\infty}O_{2l}\mathrm{e}^{\mathrm{i}l\theta},\ 
\Phi_{12}(\theta)=\sum_{l=-\infty}^{\infty}O_{2l-1}\mathrm{e}^{\mathrm{i}l\theta},\nonumber\\  
\Phi_{21}(\theta)&=&\sum_{l=-\infty}^{\infty}O_{2l+1}\mathrm{e}^{\mathrm{i}l\theta}, \ 
\Phi_{22}(\theta)=1+\sum_{l=-\infty}^{\infty}O_{2l}\mathrm{e}^{\mathrm{i}l\theta}.
\label{matrix_phi}
\end{eqnarray}
Using 
\begin{equation}
g(\phi)=\sum_{r=-\infty}^{\infty} O_r \mathrm{e}^{-\mathrm{i}r \phi}
\end{equation}
one concludes that 
\begin{eqnarray}
\sum_{l=-\infty}^{\infty}O_{2l}\mathrm{e}^{\mathrm{i}l\theta}&=&\frac{1}{2}[g(-\theta/2)+g(-\theta/2+\pi)],\nonumber\\
\sum_{l=-\infty}^{\infty}O_{2l+1}\mathrm{e}^{\mathrm{i}l\theta}&=&\frac{1}{2}[g(-\theta/2)-g(-\theta/2+\pi)]\mathrm{e}^{\mathrm{i}\theta/2},\\ 
\sum_{l=-\infty}^{\infty}O_{2l-1}\mathrm{e}^{\mathrm{i}l\theta}&=&\frac{1}{2}[g(-\theta/2)-g(-\theta/2+\pi)]\mathrm{e}^{-\mathrm{i}\theta/2}\ .\nonumber
\end{eqnarray}
Finally 
\begin{equation}
\Phi(\theta)=\left (\begin{array}{cc} -1+\tfrac{1}{2}(F(\theta)+G(\theta)) & \tfrac{1}{2} \mathrm{e}^{-\mathrm{i}\theta/2} (F(\theta)-G(\theta))\\  \tfrac{1}{2} \mathrm{e}^{\mathrm{i}\theta/2} (F(\theta)-G(\theta)) & 1+\tfrac{1}{2}(F(\theta)+G(\theta)) \end{array}   \right )
\end{equation}
where
\begin{eqnarray}
F(\theta)&=&\frac{\lambda-\cos (\theta/2) +\mathrm{i}\gamma \sin (\theta/2) }{\sqrt{(\lambda-\cos(\theta/2) )^2+\gamma^2 \sin^2  (\theta/2)}},\\ 
G(\theta)&=&\frac{\lambda+\cos (\theta/2) -\mathrm{i}\gamma \sin (\theta/2) }{\sqrt{(\lambda+\cos(\theta/2) )^2+\gamma^2 \sin^2  (\theta/2)}}\ .\nonumber
\end{eqnarray}
The above formulae give
\begin{equation}
\det \Phi(\theta)= F(\theta)G(\theta)-1
\end{equation}
The imaginary part of the $\ln \det \Phi(\theta)$ will give zero after the integration over $\theta$ and the real part is 
\begin{equation}
\mathrm{Re} \Big (\ln \det \Phi(\theta)\Big )=
\frac{1}{2}\ln \Big [ 2\Big ( 1- \frac{\lambda^2+\gamma^2 -(1+\gamma^2)\cos^2(\theta/2)}{\sqrt{R_{+}(\lambda, \gamma,\theta/2)R_{-}(\lambda, \gamma,\theta/2)} }\Big )\Big ]
\end{equation} 
where $R_{\pm}(\lambda, \gamma,u)=(\lambda\pm \cos u)^2 +\gamma^2 \sin^2 u$.

Consequently, the contribution of the N\'eel configuration to the fractal dimension is
\begin{equation}
D_{\mathrm{Neel}}(\lambda, \gamma)=
\frac{3}{4}-\frac{1}{2\pi \ln 2}\int_0^{\pi/2}\ln \Big [1- \frac{\lambda^2+\gamma^2-(1+\gamma^2)\cos^2 u}{\sqrt{R_{+}(\lambda, \gamma,u)R_{-}(\lambda, \gamma,u)}}\Big ]\mathrm{d}u\ . 
\label{D_Neel_XY}
\end{equation}
When $\lambda=0$
\begin{equation}
D_{\mathrm{Neel}}(0, \gamma)=1+\frac{1}{2\ln 2}\ln \frac{1+\gamma}{2}
\end{equation}
from which it is clear that the N\'eel configuration dominates $D_{-\infty}$ at small $\lambda$ when $\gamma>1$.

In general, if $D_{\mathrm{Neel}}>D_{-}$, $D_{-\infty}=D_{\mathrm{Neel}}$, otherwise  $D_{-\infty}=D_{-}$. For $\gamma=1.4$ these curves intersect at $\lambda\approx 0.4982$ and $D_{-\infty}$ is built from two curves. Numerical results presented in \cite{ab} agree well with this prediction.  


\section{XXZ and XYZ models} \label{XXZ}

In previous Sections we discussed the QIM and the XY model with ferromagnetic GS. Here we briefly  consider the XXZ and XYZ models choosing the parameters such that their GS are anti-ferromagnetic. 
  
The XXZ model in zero fields  is  a particular case of the Heisenberg model \eqref{general_H} with $\gamma=\lambda=\alpha=0$ and $\Delta\neq 0$
\begin{equation}
H_{XXZ}=-\sum_{n}\left(\sigma_{n}^{x}\sigma_{n+1}^{x}+ \sigma_{n}^{y}\sigma_{n+1}^{y}+\frac{\Delta}{2}\sigma_{n}^{z}\sigma_{n+1}^{z}\right)\ .
\end{equation}
Due to the conservation of the $z$ component of the total spin, $S_z=\sum_n\sigma_n^{z}$, this Hamiltonian   can be diagonalized  in subspace with fixed total spin along the $z$-axis.   The model is soluble by the coordinate Bethe anzatz \cite{bethe}, \cite{yang}, \cite{orbach} and has a rich  phase  diagram (see e.g.  \cite{mattis}). 

\begin{figure}[!ht] 
\begin{minipage}{.49\linewidth} 
\begin{center}
\includegraphics[angle=-90, width=.99\linewidth,clip]{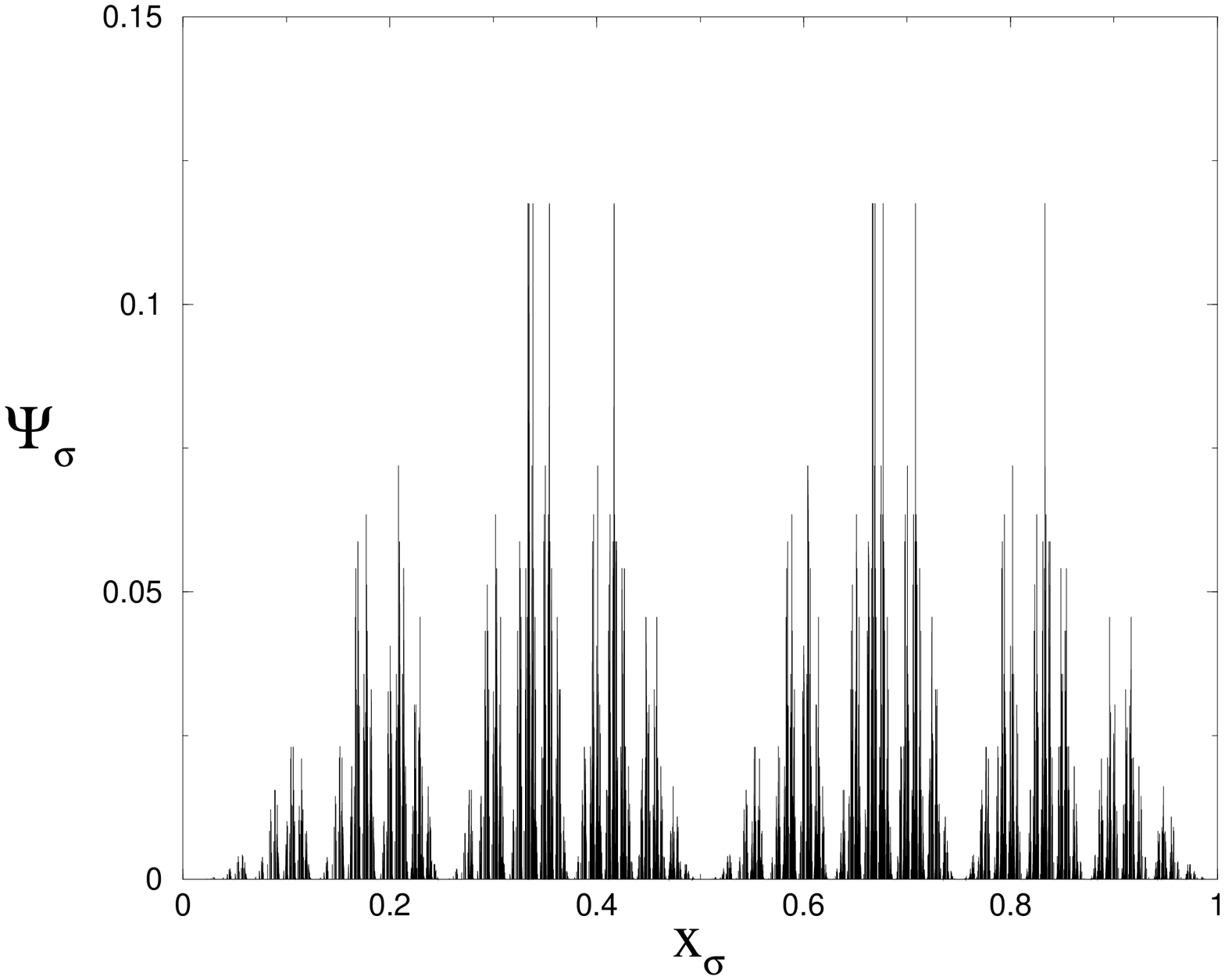}\\
(a) \end{center}
\end{minipage}
\begin{minipage}{.49\linewidth} 
\begin{center}
\includegraphics[angle=-90, width=.99\linewidth ,clip]{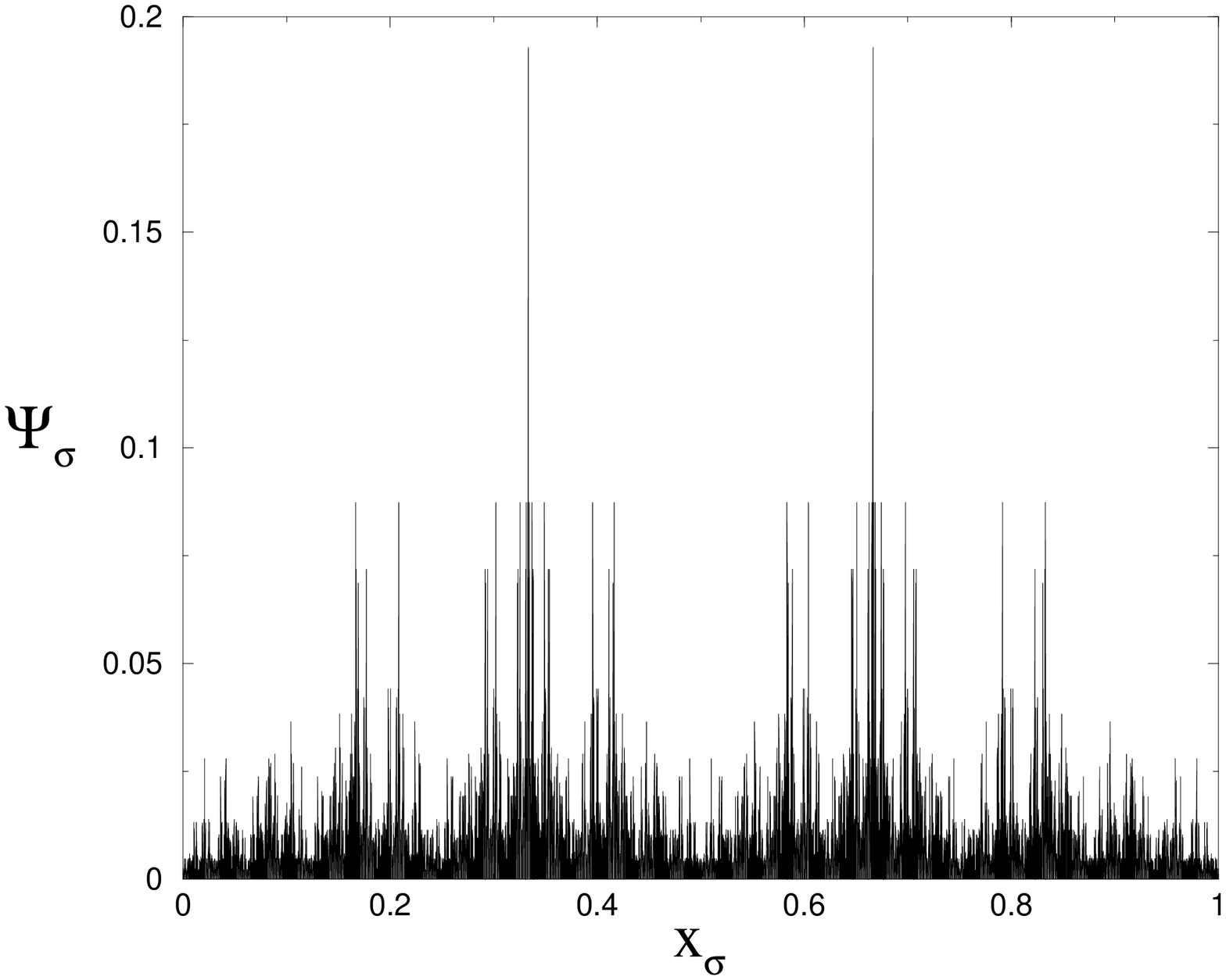}\\
(b) \end{center}
\end{minipage}
\caption{(a) Wave function of the XXZ model at  $\Delta=-\frac{1}{2}$ and $N=13$. (b) Wave function for the XYZ model in zero field with $\Delta=-0.5$, $\gamma=0.6$ and $N=12$. }
\label{wave_function_XXZ}
\end{figure}

For this model there exists a special point, $\Delta=-\tfrac{1}{2}$,  called the combinatorial point where more information about GS wave function is available.  From the Razumov--Stroganov conjecture \cite{razumov} proved in \cite{cantini} it follows that at this value of $\Delta$  and odd $N=2R+1$ the following statements are valid
\begin{itemize}
\item the GS energy is $-3N/4$.
\item The largest coefficient in the normalized ground state expansion \eqref{psi_expansion} (the one for the N\'eel configuration)  is 
\begin{equation}
\Psi_{\mathrm{max}}^{-1}=\frac{3^{N/2}}{2^N} \frac{2\cdot 5\ldots (3N-1)}{1\cdot 3\ldots (2N-1)}\ .
\end{equation}
\item The sum of all terms
\begin{equation}
\sum_{\vec{\sigma}}\Psi_{\vec{\sigma}}=3^{N/2}\ .
\end{equation}
\item The minimal coefficient corresponds to a half consecutive spins up and other spins down 
\begin{equation}
\Psi_{\mathrm{min}}^{-1}=A_R
\end{equation}
where $A_R$ is the number of alternating sign $R\times R$ matrices given by formula \cite{asm} 
\begin{equation}
A_R=\prod_{j=0}^{R-1}\frac{(3j+1)!}{(n+j)!}
\end{equation}
\end{itemize}
These formulas imply that for $\Delta=-\tfrac{1}{2}$ fractal dimensions $D_{\infty}$ and $D_{1/2}$ can be calculated analytically
\begin{equation}
D_{\infty}=\frac{3\ln 3}{2\ln 2}-2 \approx 0.3774,\quad D_{1/2}=\frac{\ln 3}{2\ln 2}\approx 0.7925\, .
\label{razumov_points}
\end{equation}
The value of the minimal coefficient determines the limiting behaviour of R\'enyi entropy \eqref{renyi_entropy}  at large negative $q$.  The asymptotics of $A_R$ when $R\to\infty$ is :  $\ln A_R= R^2\ln (3\sqrt{3}/4)  +\mathcal{O}(R)$. Hence, the smallest coefficient in the XXZ model with $\Delta=-1/2$ dereases exponentially not with $N$ but with $N^2$. This quadratic  behaviour is a particular case of the emptiness formation probability of a string of $n$ aligned spins with $n\sim N$ \cite{emptiness}. Such asymptotic behaviour  means that negative moments of the GS wave function  in anti-ferromagnetic case  require a scaling different from \eqref{fractal} and will not be considered here.  In Fig.~\ref{wave_function_XXZ} a) the wave function of the XXZ model at the combinatorial point is presented.

\begin{figure}[!ht]  
\begin{center}
\includegraphics[width=.5\linewidth, clip]{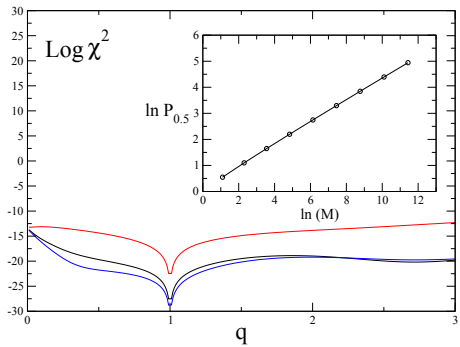} 
\end{center}
\caption{ (Online version in colour.) Plot of the logarithm of the chi-square \eqref{chi_2} as a function of  $q$ for the XXZ model in the combinatorial point $\Delta=-1/2$ for 3 different fits in \eqref{fit_h}. Red line (upper curve) corresponds to the fit $f_1$. Black line (middle curve) is obtained by the fit $f_2$. Blue line (lower curve) is for the fit $f_3$. Inset: $\ln P_{1/2}$ as a function of  $\ln M$ where $M=C_N^{[N/2]}$ with odd $N=3,\ldots,19$.}   
\label{moment_Raz}
\end{figure}

The numerical calculations were performed by extrapolation of
the R\'enyi entropy separative for odd and even $N=3-19$ (cf. Fig.~\ref{moment_Raz}). The necessity of different formulas for odd and even $N$ is physically related to the existence of different total spin projection in the GS. For even $N$ the total GS projection along $z$-axis is zero, but for odd $N$ it is $1/2$.  

To find fractal dimensions we use 3 different kinds of fits to the available data with the same number of unknowns   
\begin{eqnarray}
f_1(z)&=&a_{0}+a_{1}z+\frac{a_{2}}{z}+\frac{a_{3}}{z^2}+\frac{a_{4}}{z^3},\nonumber \\
f_2(z)&=&a_{0}+a_{1}z+\frac{a_{2}}{z}+a_{3}\ln z+\frac{a_{4}\ln z}{z}, \label{fit_h}\\
f_3(z)&=&a_{0}+a_{1}z+\frac{a_{2}}{z}+a_{3}\ln z+\frac{a_{4}}{z^2}\ .\nonumber 
\end{eqnarray}
The point is that  the exact asymptotics of $D_{\infty}$ contains the logarithmic term which corresponds to the term $a_3$ in $f_2(z)$ and $f_3(z)$.  From Fig.~\ref{moment_Raz} it is clear that the second and third  fits give better approximation to numerical data than the first one.   With available precision, fractal dimensions for odd and even $N$ are the same but sub-leading terms in the R\'enyi entropy  \eqref{renyi_entropy} are different.  The numerical calculation of $D_{1/2}$ gives $D_{1/2}=0.7949$ which agrees well with the exact result \eqref{razumov_points}.

The XYZ model is similar to the XXZ model but with the anisotropy $\gamma$ present
\begin{equation}
H_{XYZ}=-\sum_{n} \left( \frac{1+\gamma}{2} \sigma_{n}^{x} \sigma_{n+1}^{x}+
\frac{1-\gamma}{2}\sigma_{n}^{y}\sigma_{n+1}^{y}+
\frac{\Delta}{2}\sigma_{n}^{z}\sigma_{n+1}^{z}+\lambda\sigma_{n}^{z}\right ) \ .
\end{equation}
Its GS wave function can be found by the algebraic Bethe anzatz \cite{baxter}.   

This model also has a special value of the  field,
\begin{equation}
 \lambda_f=\sqrt{(1-\Delta)^2-\gamma^2}
\end{equation} 
 where the GS wave function has the factorised  form  \eqref{factorized} with 
\begin{equation}
\cos^2 2\theta=\frac{1-\gamma-\Delta}{1+\gamma-\Delta}\ .
\label{factor_angle_XYZ}
\end{equation}
As for the XY model at this field all fractal dimensions are the same as for the binomial measure \eqref{xy_exact}.

The combinatorial point for the XYZ model at zero field is $\Delta=(\gamma^2-1)/2$. At this point certain properties of the GS is known (or conjectured) \cite{rasumov_2} but they do not permit to find fractal dimensions analytically. 

As an example, we present at Fig.~\ref{wave_function_XXZ} b) the GS wave function for the XYZ model with $\Delta=-0.5$, $\gamma=0.6$ and $N=12$. Qualitatively, fractal dimensions in the XYZ model are similar to the XXZ model and are presented in \cite{ab}. 
\section{Conclusion}\label{conclusion}

Wave functions for $N$-spin problems can be represented only as a collection of exponentially large (as $N\to\infty$) number of coefficients corresponding to all possible basis states. This proliferation makes difficult not only the analysis but even the representation of such functions. There exists no 'natural' ordering of wave function coefficients and simple attempts like the use of the binary codes as in examples above lead to irregular and complex structures  even in one-dimensional models. 

The mutifractal formalism has been developed to measure quantitatively how irregular different objects are. The main result of this paper and of Ref.~\cite{ab} is the demonstration that multifractal formalism is an adequate and useful language to describe GS wave functions of spin chains. This formalism can informally be considered as an analogue of the usual thermodynamic formalism but applied not to true energies of interacting particles but to 'pseudo-energies' associated with wave function coefficients (cf. \eqref{thermodynamics}).

We stress that the models considered have no random parameters and the origin of their multifractality is not an interplay between the localization in random relief and the spreading as in critical models like the 3-dimensional Anderson at the metal-insulator point but  the complexity of internal structure of many-body Hilbert space. In this aspect, the simple example of the binomial measure is a characteristic one. It clearly show how the necessity of choosing one path from an exponentially large number of other possibilities results in irregular multifractal structure. 

The central object of the multifractal formalism is fractal dimensions, $D_q$, which play the role similar to the free energy in the usual thermodynamics with $q$ being the inverse temperature. It is well established that for classical Hamiltonians with local interactions the free energy exists in thermodynamics limit $N\to\infty$. Much less is known rigorously about fractal dimensions. Our results strongly suggest that for quantum $N$-body Hamiltonians with local interactions they do exit (i.e. the limit $N\to\infty$ give a well defined answer) but we are unaware of formal proof even in simple models like QIM.   

Only in very rare models (like spin chains in factorising field) all fractal dimensions can be calculated analytically. In some models one can find exact expression for fractal dimensions at certain values of $q$. Here we present details of the calculations of  limiting values $D_{\pm\infty}$ and $D_{1/2}$ in QIM, XY model, and XXZ model at $\Delta=-1/2$. These results are important as they prove the existence of fractal dimensions at least at special values of $q$ and can be use to control the precision of numerical calculations. 

In general fractal dimensions have to be estimated numerically and the main question is how careful the limit $N\to\infty$ can be found from the numerical data with $N$ of the order of $10-20$. In all models discussed in this paper it appears that the correction to the limit are reasonably small but the form of interpolation may depend on  $q$. In many cases the precision of $10^{-2}-10^{-3}$ can be achieved without considerable numerical efforts. 

Though in the paper only spin chain models are considered, preliminary results on one-dimensional bosonic and fermionic models demonstrate that their ground state wave functions are also  multifractals.  Based on these calculations we conjecture that the multifractality of the ground state wave function is a common feature of generic $N$-body quantum Hamiltonians with local interactions.  The discussion of multifractal properties of excited states is postponed to a future publication.
       
\ack{The authors are greatly indebted to G. Roux for many useful discussions and especially for the permission to use his code to implement the Lanczos algorithm. The numerous conversations with O. Giraud are thankfully appreciated. One of the authors (Y.Y.A) is supported by the CFM Foundation.}

\end{document}